\documentclass[preprint]{aastex}
\usepackage{rotating}
\usepackage{multirow}
\usepackage{subfigure}

\slugcomment{ApJ. in press}

\shorttitle{Multi-phase cold gas kinematic of M51}
\shortauthors{Colombo et al.}

\begin{document}

\title{The PdBI Arcsecond Whirlpool Survey (PAWS): \\
    Multi-phase cold gas kinematic of M51\footnote{Based on
observations carried out with the IRAM Plateau de Bure Interferometer and 30m telescope. IRAM is
operated by INSY/CNRS (France), MPG (Germany) and IGN (Spain).}}

\author{Dario Colombo\altaffilmark{1}, Sharon E. Meidt\altaffilmark{1}, Eva
Schinnerer\altaffilmark{1}, Santiago
Garc\'{i}a-Burillo\altaffilmark{2}, Annie
Hughes\altaffilmark{1}, J\'er\^ome
Pety\altaffilmark{3,}\altaffilmark{4}, Adam K. Leroy\altaffilmark{5},  Clare L.
Dobbs\altaffilmark{6}, 
Ga\"{e}lle Dumas\altaffilmark{3}, Todd A. Thompson\altaffilmark{7,}\altaffilmark{8}, Karl
F. Schuster\altaffilmark{3} and Carsten Kramer\altaffilmark{9}.}

\altaffiltext{1}{Max Planck Institute for Astronomy, K\"onigstuhl 17, 69117 Heidelberg, Germany}
\altaffiltext{2}{Observatorio Astron\'{o}mico Nacional - OAN, Observatorio de Madrid Alfonso XII, 3,
28014 - Madrid, Spain}
\altaffiltext{3}{Institut de Radioastronomie Millim\'etrique, 300 Rue de la Piscine, F-38406 Saint
Martin d'H\`eres, France}
\altaffiltext{4}{Observatoire de Paris, 61 Avenue de l'Observatoire, F-75014 Paris, France.}
\altaffiltext{5}{National Radio Astronomy Observatory, 520 Edgemont Road, Charlottesville, VA
22903, USA}
\altaffiltext{6}{School of Physics and Astronomy, University of Exeter, Stocker Road, Exeter EX4
4QL, UK}
\altaffiltext{7}{Department of Astronomy, The Ohio State University, 140 W. 18th Ave., Columbus,
OH 43210, USA} 
\altaffiltext{8}{Center for Cosmology and AstroParticle Physics, The Ohio State University, 191 W.
Woodruff Ave., Columbus, OH 43210, USA}
\altaffiltext{9}{Instituto Radioastronom\'{i}a Milim\'{e}trica, Av. Divina Pastora 7, Nucleo
Central, 18012 Granada, Spain}

\begin{abstract}
The kinematic complexity and the favorable position of M51 on the sky make this galaxy an ideal
target to test different theories of spiral arm dynamics. Taking advantage of the
new high resolution PdBI Arcsecond Whirlpool Survey (PAWS) data, we undertake a
detailed kinematic study of M51 to characterize and quantify the origin and nature of the
non-circular
motions. Using a tilted-ring analysis supported by several other archival datasets we update 
the estimation of M51's position angle ($PA=(173\pm3)^{\circ}$) and
inclination ($i=(22\pm5)^{\circ}$). Harmonic decomposition of the high resolution ($\sim40$ pc) CO
velocity
field shows the first kinematic evidence of an $m=3$ wave in the inner disk of M51 with a corotation
at $R_{CR,m=3}=1.1\pm0.1$ kpc and a pattern speed of $\Omega_{p,m=3}\approx140$ km s$^{-1}$
kpc$^{-1}$.
This mode seems to be excited by the nuclear bar, while the beat frequencies generated by the
coupling between the $m=3$ mode and the main spiral structure confirm its density-wave nature. We
observe
also a signature of an $m=1$ mode that is likely responsible
for the lopsidedness of M51 at small and large radii. We provide a simple method to estimate the
radial
variation of the amplitude of the spiral perturbation ($V_{sp}$) attributed to the different modes.
The main spiral arm structure has $\langle V_{sp}\rangle=50-70$ km s$^{-1}$, while the streaming
velocity associated with the $m=1$ and $m=3$ modes is, in general, 2 times lower. Our joint
analysis
of HI and CO velocity fields at low and high spatial resolution reveals that
the atomic and molecular gas phases respond differently to the spiral perturbation due to their
different vertical distribution and emission morphology.     
\end{abstract}

\section{Introduction}\label{sec:intro}
\noindent Gas kinematics are key to dissecting how the various components of a galaxy (stars, gas and dust)
interact and evolve over time,  leading to the variety of morphologies we see in the local
universe today. They supply the standard for probing the mass distributions of galaxies through
rotation
curves and are uniquely sensitive to perturbations to the gravitational potential due to bars and
spiral arms (\citealt{roberts87}; \citealt{vogel93}; \citealt{regan01}; \citealt{dobbs10}). By
providing an
instantaneous record of the response of gas to non-axisymmetric (bar and spiral) structures, they
supply a unique view of the processes by which these features impact the distribution of gas and
stars, from stimulating stellar radial migration (\citealt{sellwood02}; \citealt{minchev12}) and
driving gas
inflows (\citealt{wong04}; \citealt{vandeven10}) to regulating the conversion of gas into stars
(\citealt{meidt13}).    Gas kinematics are therefore indispensable for building a firm picture of
how bar and spiral structures contribute to 
the slow, secular evolution of galaxies.\\  

\noindent Studying the response of gas to an underlying potential perturbation (in the form of bars or spiral
arms) can supply key information about the nature of the perturbation (e.g. \citealt{vogel93};
\citealt{wong04}). 
Today, spiral structures tend to be described by one of two opposite theories. 
In the
quasi-stationary spiral structure (QSSS) depiction (\citealt{lindblad63}), spiral arms are a long
lasting
pattern (\citealt{lin64}) that slowly evolves and rotates with a single angular speed. This
structure is thought to be
formed from self-excited
and self-regulated standing ``density waves'' (\citealt{bertin89a}; \citealt{bertin89b}; \citealt{bertin96}) present in the density and hence gravitational potential.  The
other theory considers arms to be transient disturbances generated, e.g., by
the tidal interaction with a companion (e.g., \citealt{toomre72}) which overwhelms any
pre-existing
structure (\citealt{salo00}) or given some initial seed perturbation (\citealt{donghia13}). 
These structures, which may not obey the Lin-Shu dispersion relation for density
waves (\citealt{salo00}, \citealt{donghia13}), are often thought to be winding (with
radially decreasing pattern speeds) or to consist of material moving at series of distinct speeds.\\

\noindent Most of the effort to discriminate between these two theories has been centered on M51, which is an
ideal target because of its proximity (D=7.6 Mpc, \citealt{ciardullo02}), favorable inclination
($i\sim22^{\circ}$, this work), high surface brightness and kinematic complexity.  
In the seminal M51 kinematic study of \cite{tully74c}, the spiral pattern in the outer disk was
identified as a transient feature stimulated by the interaction between M51a and M51b, while the inner arms
were thought to be in a steady state.  Indeed, \cite{vogel93} find very good agreement between
the predictions of density-wave theory and the observed transverse velocities across the inner
arms. But more recently, \cite{shetty07} argue that gas density and velocity profiles are
inconsistent with quasi-steady state mass conservation.\\  

\noindent At least some of the ambiguity regarding the nature of M51's spiral pattern may stem from the
complexity of its structure.  \cite{meidt08} found evidence for three distinct pattern speeds
in M51 using the radial Tremaine-Weinberg (TWR) method, only one of which is similar to the value
typically assumed.  Their finding that these patterns overlap at resonances would seem to be
consistent with the idea that they are physically coupled and not temporary disturbances.  But
multiple, distinct pattern speeds may also support the \cite{donghia13} picture wherein
a disturbance drives a transient feature that stimulates other transient features, which together
give the appearance of long-lived structures.\\  

\noindent The disk of M51 may also sustain multiple, spatially coincident patterns.  The optical and NIR
surface brightness is clearly lopsided, suggesting an $m=1$ disturbance in the potential.  This
lopsidedness persists in tracers of the ISM.  Some part of the lopsidedness could be explained by
the superposition of the two-armed spiral with a spiral pattern with three-fold symmetry
(\citealt{henry03}).   
The existence of such a pattern in M51 was first suggested between radii of 50" and 100"  in blue
light
optical images by \cite{elmegreen92}.  \cite{rix93} also find the
signature of a three-armed pattern in the K-band, although at a much weaker level than in the
V-band.\footnote{They also found that M51a is lopsided at all radii, as indicated by the high power
in the $m=1$ Fourier component.}  Both studies conclude that the $m=3$ feature in M51 is a
perturbation in
the gas and dust only (traced in extinction at optical and NIR wavelengths), rather than a genuine
density wave present in the density (traced by the 
old stellar light) and thus gravitational potential of the system, although this idea was later
challenged by \cite{henry03}. As pointed out by \cite{elmegreen92}, simple Fourier transforms of galaxy
images can provide misleading results on the nature and number of spiral arms if they are not
confirmed by kinematic evidence.  The $m$=3 component, for example, could arise as a beat frequency,
modulated by inter-arm star formation or by an intensity gradient from one side of the galaxy to the
other (due to extinction or kinematic effects).\\

\noindent In this paper we take advantage of the new high resolution $^{12}$CO\,(1-0) PAWS
observations in the central 9 kpc of M51.  The high resolution of this data ($\sim1"$) allows us to
perform an in-depth study of the gas response to M51's perturbed stellar potential.   If the $m=3$
mode is a genuine perturbation to the potential then our high resolution map of molecular gas
motions should reveal it. We complement our kinematic analysis with lower resolution HI and $^{12}$CO\,(2-1) data
from THINGS (\citealt{walter08}) and HERACLES (\citealt{schuster04}, \citealt{leroy09}). The
inclusion of observations of various phases of the ISM, at low and high resolution, allows us to
assess how uniformly they trace the gravitational potential, and determine which type of observational
tracer is optimal for which science goal.  The 21 cm and the CO line emission are the common tracers of the
atomic and the molecular gas phases that are at the basis of star formation.  To understand the
physics behind empirical laws that relate gas and stars from kpc (e.g. \citealt{leroy13},
\citealt{bigiel08} 
and references therein) to pc scales it is necessary to constrain their characteristics at
every level, especially how they are distributed within, and respond to the potential of, a given
system.  \\

\noindent The paper is constructed in the following way. In Section~\ref{sec:data} we present the
datasets used for our kinematic analysis. Then we describe the features of the high resolution
velocity field from PAWS in Section~\ref{sec:mom_descr_mom1}. We introduce the formalism to study
the
line-of-sight velocity ($V_{los}$) in spiral galaxies in Section~\ref{sec:vlos} together with our
estimation of the projection parameters of M51 (inclination and position angle) needed for a correct
evaluation of the single component of $V_{los}$. 
In Section~\ref{sec:non_circ} we use the harmonic decomposition
prescriptions to study residual velocity fields. We propose a method to estimate the amplitude of
the perturbation velocity from the spiral arms and we present the first kinematic evidence for a
three-fold density-wave in M51. We conclude in Section~\ref{sec:disc_m3} discussing the origin of this
structure and highlighting kinematic differences between atomic and molecular gas tracers and low
and high resolution data (Section~\ref{sec:disc_dataset}). We summarize our work and findings in Section~\ref{sec:summary}.\\

\section{Data}\label{sec:data}

\subsection{PAWS $^{12}$CO(1-0) data}
\noindent The PdBI Arcsecond Whirlpool Survey (PAWS, \citealt{schinnerer13}) ``hybrid cube'' considered here has been obtained by combining the IRAM-30m single-dish antenna and Plateau de Bure Interferometer (PdBI) $^{12}$CO(1-0) observations of M51 (\citealt{pety13}). The cube has an angular resolution of $1''.16\times0''.97$ (or $\sim40$\,pc at $7.6$\,Mpc distance, \citealt{ciardullo02}), a mean RMS noise of $\sim 0.4$\,K per 5 km\,s$^{-1}$ channel and covers the LSR velocity range between 173 to 769 km s$^{-1}$. PdBI dedicated observations of the inner disk of M51a (field-of-view, FoV $\sim270''\times170''$ or $\sim11\times6$\,kpc) were carried out in the A, B, C, and D configurations from August 2009 and March 2010.\\

\noindent We also independently consider the 30m single-dish observations (hereafter indicated with the name
\emph{30m}) of the full disk of M51 
($\sim60$ square arcminutes) conducted to recover the low spatial frequency information filtered out by the PdBI. This data has a spatial resolution of $22.5"$ (i.e. $\sim900$\,pc at 7.6\,Mpc distance) and a channel width of $\sim5$ km\,s$^{-1}$.\\

\noindent To study the impact of resolution, we also include the hybrid data cubes gaussian-tapered to a synthesized
resolution of 3" and 6" presented in Pety et al. (2013) with typical RMS noise 
of 0.1 and 0.03 K, respectively. These PAWS datasets span the same range of
LSR velocities and have the same field-of-view as the PAWS dataset at 1".\\

\subsection{Archival THINGS VLA HI data}
\noindent M51 HI data from The HI Nearby Galaxy Survey (THINGS, Walter et al. 2008) was obtained
from the dedicated web-page \verb"http://www.mpia-hd.mpg.de/THINGS/Data.html". M51 was observed
between March 2005 and July 2007 using the NRAO Very Large Array (VLA) in B, C and D
configuration. The robust-weighted THINGS data used here has a spatial resolution of $\sim6"$ (i.e.
240 pc at our assumed
M51 distance of 7.6\,Mpc) and a spectral resolution of $\sim5$ km\,s$^{-1}$.
The $1\sigma_{RMS}$ noise sensitivity of the survey is homogeneous and $\sim6$\,K per channel. We use this data
together with the PAWS data to better define the rotation curve of M51, as it covers the entire disk
of M51a.\\

\subsection{Archival HERACLES IRAM 30m $^{12}$CO(2-1) data}
\noindent The Heterodyne Receiver Array CO Line Extragalactic Survey (HERACLES, \citealt{leroy09})
re-reduced and mapped the data previously obtained for M51 by \cite{schuster04} using the HERA receiver array on the
IRAM 30m telescope from January
2006 through March 2008 for M51. The $^{12}$CO\,(2-1) M51 data has a spatial resolution of
$\sim13.5"$ (540 pc in M51) and
a spectral resolution of $\sim2.6$ km\,s$^{-1}$. M51 data presents an $1\sigma_{RMS}$ noise sensitivity $\sim22$\,mK per channel.\\

\section{M51a neutral gas velocity fields}\label{sec:mom_descr_mom1}
\begin{figure}[h]
\begin{center}
\includegraphics[width=1 \textwidth]{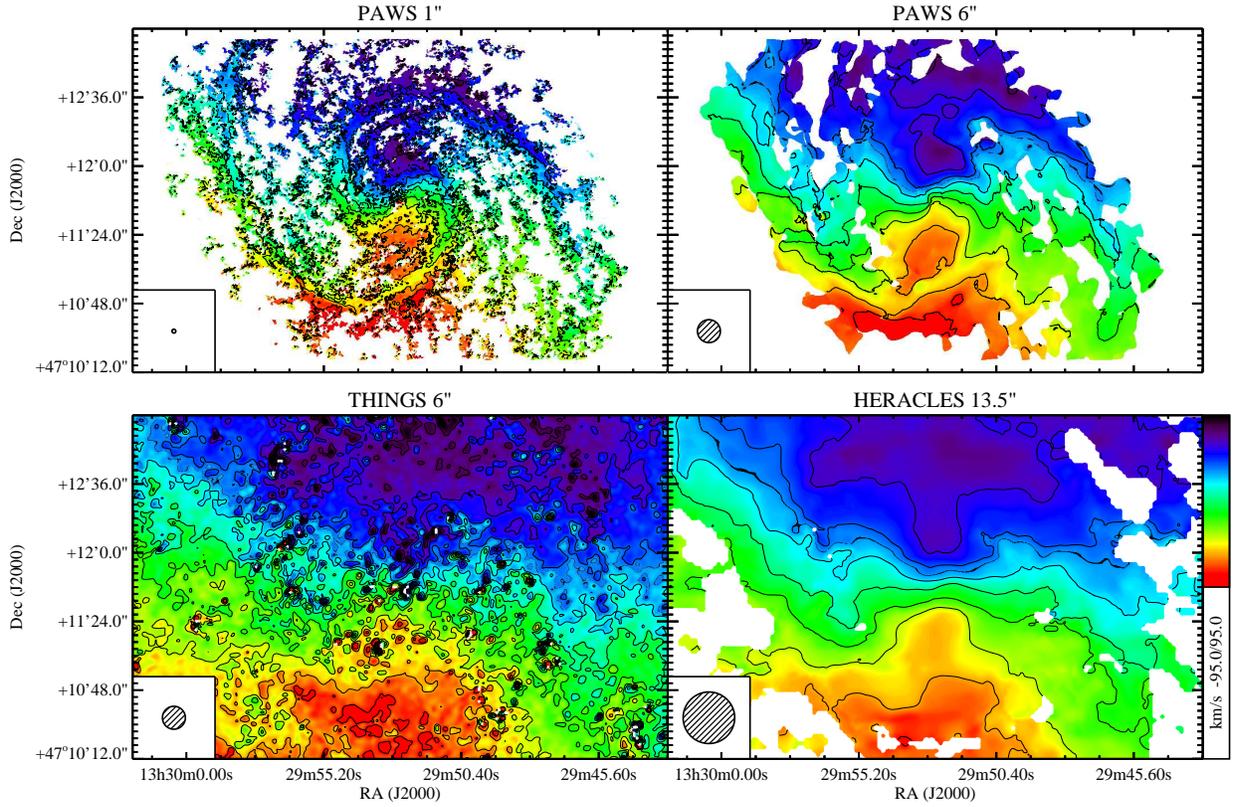} 	
\end{center}
\caption{\footnotesize The PAWS $^{12}$CO(1-0) Velocity field at 1" resolution (top left). Deviations from circular motion are due to streaming motions associated with the spiral arms and the nuclear stellar bar which causes the twist in the inner line-of-nodes.  M51's nucleus is also redshifted with respect
to the systemic velocity of the galaxy. These features are progressively smeared out by the beam in the  PAWS tapered 6" (top right), THINGS 6" (bottom left) and HERACLES 13.5" (bottom right) velocity fields. The sidebar shows the color scale of the map in km\,s$^{-1}$ relative to the systemic velocity of M51, 472 km\,s$^{-1}$ (\citealt{shetty07}). In the bottom left of each panel the beam is indicated (where 1"$\sim40$\,pc).}
\label{fig:mom1_other}
\end{figure}

\noindent In the following we will utilize the moment maps (velocity field, velocity dispersion map) derived for our different neutral gas cubes following the masking method described in the Appendix B of \cite{pety13}. The PAWS 1" velocity field (top left of Fig \ref{fig:mom1_other}) exhibits significant deviations from pure circular motion (visible in the irregularity of line-of-nodes), the most prominent of which are: strong spiral arm streaming motions, a twist in the central region and the nucleus of M51a itself.\\

\noindent The streaming motions associated with the spiral arms are particularly evident in the
southern half of the PAWS FoV, characterized by discontinuities and velocity gradients across the arm.  The deviation persists to a much lesser degree in parts of the inter-arm region. Streaming motions appears less strong in the northern compared to the southern half.\\

\noindent In the central region ($R\lesssim35"$) the iso-velocity contours are strongly twisted
by $10-15^{\circ}$. A recent torque analysis (\citealt{meidt13}) suggests that the observed twisting is due to the nuclear bar
first seen in near-IR images (\citealt{zaritsky93}).  At the very center of the map, the nuclear gas shows a
clear out-of-velocity pattern redshifted by $\approx100$ km
s$^{-1}$ with respect to the systemic velocity (see also \citealt{scoville98}, \citealt{matsushita07}).\\

\noindent The prominence of these features is reduced at degraded resolution, as they are largely smeared out by a larger beam. To illustrate this, in Fig.~\ref{fig:mom1_other} we show the first moment maps from PAWS
tapered at 6", THINGS at 6" and HERACLES at 13.5". In PAWS 6" the
redshifted nucleus is not visible and the discontinuities of the velocity gradient
across the arms are
strongly reduced. These features are completely absent in the THINGS and HERACLES first moment maps.
While in the case of HERACLES this absence could be due to the much lower resolution and the lack
of interferometric data, the difference between the CO and HI data at the same resolution could be due
to a real difference in the nature and distribution of the two emission line tracers. We discuss this possibility in
Section~\ref{sec:disc_dataset}.\\

\clearpage
\newpage

\clearpage
\newpage

\section{Gas motions in spiral potentials}\label{sec:vlos}
\noindent In this section and the next, we consider the different velocity components that contribute along
the line-of-sight in a typical spiral galaxy in the presence of strong non-circular motions.  Each
component is analyzed in detail in order to gain an optimal view of cold gas kinematics in M51, as
well as to explore how this view depends on the resolution at which the gas motions are observed.\\  

\subsection{Line-of-sight velocity}
\noindent The line-of-sight
velocity $V_{los}$ observed at a given location in a galactic disk can be represented as a sum of
four
parts:

\begin{equation}\label{eq:vlos}
 V_{los}=V_{sys}+V_{rot}+V_{pec}+V_{z}
\end{equation}

\noindent where $V_{sys}$ is the systemic velocity of the galaxy due to the expansion of the
Universe, $V_{rot}$ is the rotational component, $V_{pec}$ represents all peculiar velocities not
accounted for the circular motion of the galaxy and $V_{z}$ is the vertical velocity component (i.e.
\citealt{canzian97}). Studies of face-on grand-design spirals indicate that $V_z$ of the neutral
gas 
is less than 5 km s$^{-1}$ (\citealt{vanderkruit82}), in which case $V_{los}$ can be well
represented by planar motion without considerable vertical motions. Therefore throughout this
paper we assume $V_{z}\equiv0$.\\ 

\noindent The \emph{rotational component} can be expressed as

\begin{equation}\label{eq:vrot}
 V_{rot}=V_{c}\cos(\theta)\sin i,
\end{equation}

\noindent where $V_{c}$ is the circular rotation speed, $\theta$ is the angle in the plane of the
disk from the major-axis receding side, and $i$ represents the inclination of the disk to the plane
of the sky. (The inclination $i$ is equal to $0^{\circ}$ for an exactly face-on galaxy and
$i=90^{\circ}$ for a completely edge-on geometry.)\\

\noindent In a grand-design spiral galaxy such as M51, the \emph{peculiar component} is largely due to the gas
response to the density wave perturbation, i.e. 
\begin{equation}\label{eq:vpec1}
 V_{pec}=(u_{\phi}\cos\theta+u_{r}\sin\theta)\sin i
\end{equation}

\noindent where $u_{r}$ and $u_{\psi}$ are the (non-circular) radial and azimuthal components of
streaming motions.

\subsection{Kinematic parameter estimation}\label{sec:kinpar}
\noindent Our main goal in this paper is to measure and analyze the streaming motions in the inner disk of
M51.  To correctly interpret the line-of-sight projections of peculiar motions
(i.e. $V_{pec}$) we must therefore first have a good knowledge of the kinematic parameters that describe
the projection of the galaxy on the plane of the sky.  Several parameters are already well-constrained
in the literature and do not
require further analysis (Section~\ref{sec:m51_studies}).  For others, we provide new estimations --
with uncertainties (Section~\ref{sec:tilt_ring}) -- applying a tilted-ring analysis to the different
velocity fields from PAWS 1", PAWS 3", THINGS 6", HERACLES 13.5" and 30m at 22.5".\\
   
\subsubsection{Previous M51 kinematic studies}\label{sec:m51_studies}
\noindent Because of its proximity, favourable inclination and prominent spiral arms,
M51 has been the focus of a large number of kinematic studies aimed at testing theories of spiral
arm formation and evolution.  A summary of those focused on the determination
of the kinematic parameters is provided in Tables \ref{tab:kin_par_old}-\ref{tab:center_old}.\\ 

\noindent In general the systemic velocity of M51 is well-constrained around a value 
$V_{sys}=472$ km\,s$^{-1}$. Therefore, in the following we adopt the literature value for this
quantity (e.g. \citealt{shetty07}).\\ 

\noindent The center of M51, corresponding to the location of the nucleus, has been carefully constrained by
measurements of H$_{2}$O maser emission and high resolution radio continuum imaging (see
Table~\ref{tab:center_old} and references therein). Throughout this paper we adopt as rotation
center the latest measurement of the water maser by \cite{hagiwara07}, i.e.
$(x_{0},y_{0})=(13^{h}29^{m}52^{s}.71,47^{\circ}11'42".79)$. The adopted rotation
center almost coincides with the peak of CO emission associated with M51a's bright core (located at
$(x_{core},y_{core})=(13^{h}29^{m}52^{s}.62,47^{\circ}11'42".58)$), clearly identifiable only by
PAWS at 1".\\

\noindent Estimates for the position angle $PA$ and inclination $i$ span a large range in the literature (see
Table~\ref{tab:kin_par_old} and references therein), between PA=$165-180^{\circ}$ and $i$=
$15-28^{\circ}$. With the aim of updating these estimates and providing a tighter constraint, in the
next section 
we apply a tilted-ring analysis to the most recent high-resolution gas velocity fields
available for M51 from the THINGS, HERACLES and PAWS projects.\\ 

\newpage

\begin{deluxetable}{cccccccc}
\tabletypesize{\scriptsize}
\setlength{\tabcolsep}{0.04in}
\tablecaption{M51a (NGC 5194) kinematic parameters measured by previous
studies\label{tab:kin_par_old}}
\tablehead{
\colhead{\textbf{Resolution}} &
\colhead{\textbf{Tracer}} &
\colhead{$V_{sys}$} &
\colhead{$PA$} & \colhead{$i$} &
\colhead{\textbf{Reference}} \\
}
\tablewidth{0pt}
\startdata
2"/4" & H$\alpha$/$^{12}$CO(1-0) & $471.7\pm0.3$ &$175\pm5$ & $24\pm3$ & (1)\\
4" & $^{12}$CO(1-0) & $469$ &$170\pm5$ & ... & (2)\\
5" & H$\alpha$ & $470\pm2$ & ... & ... & (3)\\
6" & HI & ... &... & 30 & (4)\\
6".75 & H$\alpha$ & $472\pm3$ & $170\pm3$ & $20\pm5$ & (5)\\
16" & $^{12}$CO(1-0) & $469\pm5$ & $171.6$ & ... & (6)\\
\enddata
\tablecomments{(1), \cite{shetty07}; (2), \cite{meidt08}; (3), \cite{goad79}; (4),
\cite{deblok08}; (5), \cite{tully74b}; (6) \cite{kuno97}. 
In \cite{shetty07} and \cite{meidt08}, 4" refers to the best resolution of the BIMA-SONG data
used.}
\end{deluxetable}

\begin{deluxetable}{cccc}
\tabletypesize{\scriptsize}
\setlength{\tabcolsep}{0.04in}
\tablecaption{Center of M51a (NGC 5194) as derived from previous studies\label{tab:center_old}}
\tablehead{
\colhead{\textbf{Resolution}} &
\colhead{\textbf{Method}} & 
\colhead{$x_{0},y_{0}$} &
\colhead{\textbf{Reference}} \\
}
\tablewidth{0pt}
\startdata
$\sim$0".1 & H$_{2}$O maser spot  & $13^{h}29^{m}52^{s}.71,47^{\circ}11'42".79$ & (1)\\
$\sim$0".1 & H$_{2}$O maser spot & $13^{h}29^{m}52^{s}.71,47^{\circ}11'42".80$ & (2)\\
 1" & 6-20\,cm continuum peak & $13^{h}29^{m}52^{s}.70,47^{\circ}11'42".60$ & (3)\\
 1".1 & 6\,cm radio continuum peak &$13^{h}29^{m}52^{s}.71,47^{\circ}11'42".61$ & (4)\\
$\sim$1".3 & 6-20\,cm continuum peak & $13^{h}29^{m}52^{s}.71,47^{\circ}11'42".73$ & (5)\\
... & Optical measurement & $13^{h}29^{m}53^{s}.27,47^{\circ}11'48".36$ & (6)\\
\enddata
\tablecomments{(1), \cite{hagiwara07}; (2), \cite{hagiwara01}; (3), \cite{ford85}; 
(4), \cite{turner94}; (5), \cite{maddox07}; (6), \cite{dressel76}. 
(B1950) coordinates reported by several studies have been converted to (J2000) using NED.}
\end{deluxetable}

\clearpage
\newpage

\subsubsection{Tilted-ring analysis}\label{sec:tilt_ring}
\noindent To quantify the kinematic parameters of M51a we assume that the various
quantities of Eq.~\ref{eq:vlos} vary only with galactocentric radius $R$. In this case, the first
moment of the line-of-side velocity distribution can be studied through a standard tilted ring
approach (\citealt{rogstad74}). We perform a least-square tilted-ring fit to the line-of-sight velocity field using the GIPSY task ROTCUR, sampling the velocity field at one radial bin per synthesized beam width from a starting radius of one half-beam.\\ 

\noindent We implement a two step procedure to obtain estimates of M51a's
kinematic parameters ($i$, $PA$):

\begin{itemize} 
 \item First we fix the systemic velocity and rotational center using the literature
values discussed in Section \ref{sec:m51_studies}, i.e.
$V_{sys}=472$ km\,s$^{-1}$ and ($x_{0},y_{0}$)=($13^{h}29^{m}52.41^{s}$, $47^{\circ}11'42.80"$), and
$V_{rad}=0$ but leaving free inclination $i$, position angle $PA$ and
rotation velocity $V_{rot}$. We estimate the magnitude of $\langle PA\rangle$ and $\langle i\rangle$
as weighted medians along the radial profile, using the inverse of the squared-errors calculated
directly by \verb"ROTCUR" as weights.  These errors are typically larger at large galactocentric
radius where the data sampling is lower. 

 \item In the second step we set different values of inclination
(i.e. $20^{\circ}$, $23^{\circ}$, $25^{\circ}$, $27^{\circ}$, $30^{\circ}$, $33^{\circ}$,
$35^{\circ}$,
$37^{\circ}$, $40^{\circ}$, $45^{\circ}$) to obtain our final position angle\footnote{$V_{sys}$ and
($x_{0},y_{0}$)
are also kept fixed as in the first step}. For every fixed inclination we calculate the weighted
median as a function of radius.
Then we apply this same procedure to obtain the inclination itself, fixing different values of $PA$
(i.e. $165^{\circ}$, $167^{\circ}$, $170^{\circ}$, $172^{\circ}$,
$173^{\circ}$, $174^{\circ}$, $175^{\circ}$, $177^{\circ}$, $180^{\circ}$, $185^{\circ}$).
\end{itemize}

\noindent The final results of the two steps are summarized in Table
\ref{tab:kin_par}. Alongside our analysis of the PAWS 1" and 3" velocity fields, we 
perform the tilted ring analysis of the 6'' THINGS HI velocity field\footnote{The original 6"
velocity field from THINGS has been cut using the GIPSY task BLOT in order to
eliminate the warped region of the outer HI disk.} (\citealt{walter08}),
the HERACLES $^{12}$CO(2-1) first moment map at 13.5" (\citealt{leroy09}) and the 30m
data at 22.5" (\citealt{pety13}).  These maps all extend beyond the PAWS field of view and allow us
to sample the full disk of M51a.  Compared to the hybrid PAWS data, these maps should also be less
sensitive to the contribution of non-circular streaming motions, which are progressively smeared out
the lower the angular resolution. 
As described in Section
\ref{sec:mom_descr_mom1}, strong spiral arm streaming motions cause distortions in the
iso-velocity
contours in the PAWS velocity field at 1" (see
Figure~1-2), which influence the estimate of the position angle. 
Tilted-ring solutions from these independent data sets with a larger field-of-view also provide a
much-needed consistency check on estimates from the PAWS data, given that the close to face-on
orientation can make it difficult to reliably assess the kinematic parameters.\\

\begin{deluxetable}{cccc}
\tabletypesize{\scriptsize}
\setlength{\tabcolsep}{0.04in}
\tablecaption{Kinematic parameters from the tilted-ring analysis\label{tab:kin_par}}
\tablehead{
\colhead{\textbf{Map}} & \colhead{\textbf{Step}}  & \colhead{$\mathbf{\langle i\rangle}$}  &
\colhead{$\mathbf{\langle PA\rangle}$}\\
 & & \colhead{$[deg]$}  & \colhead{$[deg]$}\\}
\tablewidth{0pt}
\startdata
\multicolumn{1}{c|}{\multirow{3}{*}{PAWS 1"}}&1&$48\pm7$&$177\pm4$\\
\multicolumn{1}{c|}{}&2&$45\pm8$&$177\pm4$\\
\hline
\multicolumn{1}{c|}{\multirow{3}{*}{PAWS 3"}}&1&$54\pm8$&$176\pm5$\\
\multicolumn{1}{c|}{}&2&$48\pm10$&$177\pm4$\\
\hline
\multicolumn{1}{c|}{\multirow{3}{*}{THINGS 6"}}&1&$30\pm12$&$172\pm2$\\
\multicolumn{1}{c|}{}&2&$22\pm5$&$173\pm3$\\
\hline
\multicolumn{1}{c|}{\multirow{3}{*}{HERACLES 13.5"}}&1&$30\pm6$&$171\pm4$\\
\multicolumn{1}{c|}{}&2&$25\pm7$&$172\pm4$\\
\hline
\multicolumn{1}{c|}{\multirow{3}{*}{30m 22.5"}}&1&$35\pm4$&$174\pm2$\\
\multicolumn{1}{c|}{}&2&$22\pm3$&$171\pm4$\\
\hline
\enddata 
\tablecomments{Weighted median and median absolute deviation (MAD) of kinematic parameters
(inclination $\langle i\rangle$, position angle
$\langle PA\rangle$) derived for each survey following the two steps described in the
text.}
\end{deluxetable}

\noindent In all data sets, we find that the position angle of M51a is fairly robust to changes in the assumed
inclination.  The PA is more sensitive to the presence of streaming motions, however.  While we find
$\langle PA\rangle\sim170^{\circ}-173^{\circ}$ from the low
resolution data where the influence of the streaming motion is
reduced (i.e. from 30m, HERACLES or THINGS data), the $\langle PA\rangle$ increases to
$\sim176^{\circ}$
for the PAWS data at 1" and 3" resolution.\\

\noindent Streaming motions also influence the inclination estimates, which we find to be especially
sensitive to
the assumed position angle (yielding larger error bars).  Considering that the strongest streaming
motions in M51 appear in the central 5 kpc and weaken at larger galactocentric radius (where the
outer spiral pattern is weaker), the FoV of a given survey largely determines the
value of the inclination that can be retrieved.  For maps with large FoV (30m, HERACLES and THINGS)
the inclination is low
($\langle i\rangle\sim22^{\circ}-25^{\circ}$), while for PAWS at 1" covering a smaller
FoV, the average inclination is higher than $40^{\circ}$.  We note that our tilted ring analysis
avoids the outer warp in M51 (as obvious in the HI distribution).  Since we sample the maps with
large FOVs only up to the start of the warp, our inclination and position angles are representative
of the disk.\\  

\noindent Since the THINGS HI survey has the largest FoV and probes the (outer) part of the disk where we
expect a lesser contribution from streaming motions,  
we adopt estimates from this data as our final, best measurements of the kinematic parameters: 
i.e. $\langle i\rangle=(22\pm5)^{\circ}$ and $\langle PA\rangle=(173\pm3)^{\circ}$.  These exhibit the smallest
error bars and the most constant behavior for various set values of $PA$ and $i$, respectively (Step
2). These results are consistent with the most recent measurements of the projection parameters
performed by Hu et al. 2013, ($PA=(168.0\pm2.5)^{\circ}$, $i=(20.3\pm2.8)^{\circ}$), using a
parametrization of M51's spiral arms imaged in $i-$band by the SDSS (Data Release 9). The more
constant behavior of the $PA$ and $i$ indicated by the HI compared to CO
datasets might also reflect the different natures of the atomic and molecular gas phases (see
Section~\ref{sec:disc_dataset}).\\

\section{Non-circular motions}\label{sec:non_circ}
\noindent  As is clear by a simple examination of the PAWS velocity field, gas motions in M51
deviate strongly from pure circular motion.  The non-axisymmetric stellar bar and spiral arms
drive strong radial and azimuthal ``streaming'' motions, which contribute to the term $V_{pec}$ in
Eq.~\ref{eq:vlos} and become apparent when removing a circular velocity model from the
observed velocity field.\\ 

\noindent In the following we analyze the peculiar motions that are not described by the model of
pure circular motion.  We start by summarizing the main features in the residual velocity field, obtained by subtracting a 2D projected model of the best estimate of $V_c$ from the observed velocity field. Then we describe and investigate in detail the residual velocity field and its features using a harmonic decomposition (\citealt{schoenmakers97}).
Finally, we use the results of the harmonic decomposition to estimate the amplitude of the spiral arm
streaming motions.\\

\subsection{Residual velocity fields}
\label{sec:res_vel}
\begin{figure*}[hb]
\begin{center}
\includegraphics[width=0.65\textwidth]{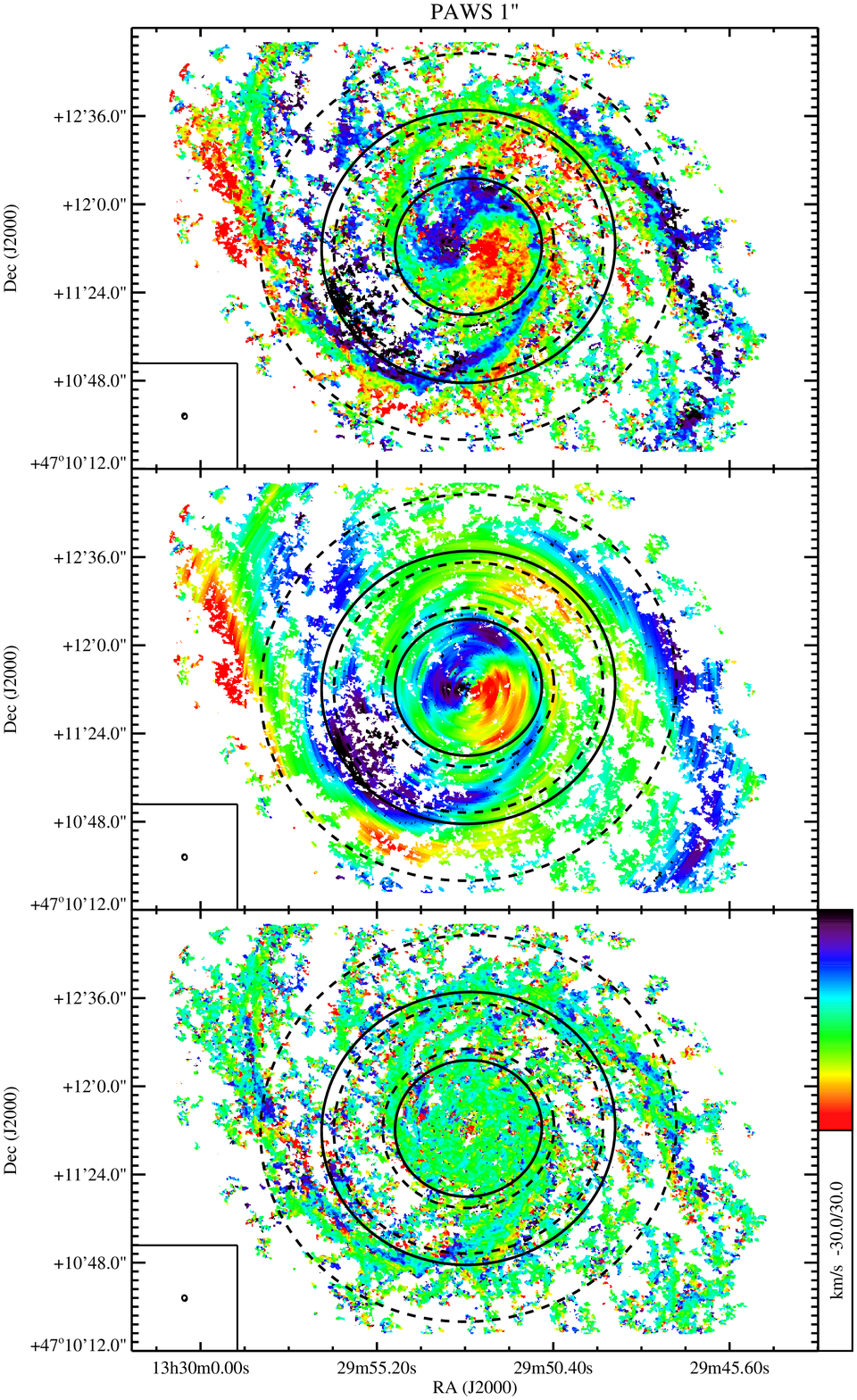}
\end{center}
\caption{\footnotesize \emph{Top}: from left to right, PAWS 1" residual velocity field.
The inner dashed black
circle indicates the outer boundary of the molecular ring
($R=35"$). The outer black dashed circles mark the
radial location of the first corotation at $R=55"$ and the material arms at $R=85"$ as identified
through the present-day torque analysis
by \cite{meidt13}. The solid black circles indicate the corotation identified
with the harmonic decomposition at $R=30"$ and $R=60"$. Individual pixels within the residual velocity fields exhibit values between  and  km\,s$^{-1}$, but we restrict the color stretch to values between -30 and +30 km\,s$^{-1}$ to highlight the main features of the residual velocity field. $\sim95\%$ of pixels have values that fall within the range [-30, 30]\,km\,s$^{-1}$. \emph{Middle}: Harmonic
reconstructed residual velocity field. \emph{Bottom}: Difference between the observed residual velocity field and its harmonic reconstruction. The beam is indicated in the bottom left of each panel.}
\label{fig:paws_res}
\end{figure*}

\begin{figure*}
\begin{center}
\includegraphics[width=0.85\textwidth]{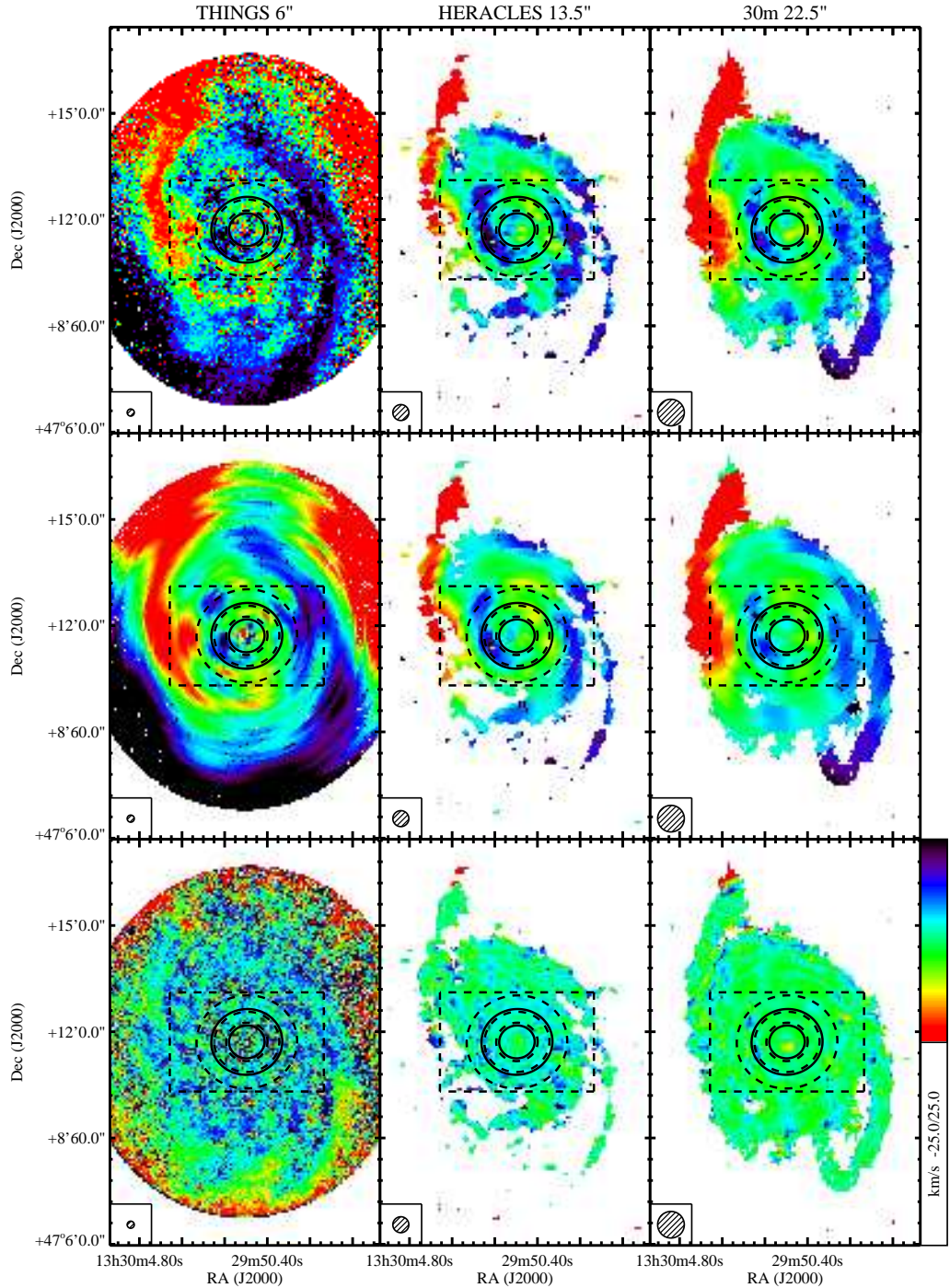}
\end{center}
\caption{\scriptsize \emph{Top}: from left to right, residual velocity fields from THINGS HI,
HERACLES $^{12}$CO(2-1)
and 30m $^{12}$CO(1-0). The inner dashed black
circle indicates the outer boundary of the molecular ring
($R=35"$). The outer black dashed circles mark the
radial location of the first corotation at $R=55"$ and the material arms at $R=85"$ as identified
through the present-day torque analysis
by \cite{meidt13}. The solid black circles indicate the corotation identified
with the harmonic decomposition at $R=30"$ and $R=60"$. Individual pixels within the residual velocity fields exhibit values between  and  km\,s$^{-1}$, but we restrict the color stretch to values between -25 and +25 km\,s$^{-1}$ to highlight the main features of the residual velocity field. $\sim95\%$ of pixels have values that fall within the range [-25, 25]\,km\,s$^{-1}$. \emph{Middle}: Harmonic
reconstructed residual velocity field. \emph{Bottom}: Difference between the observed residual velocity field and its harmonic reconstruction. The beam is indicated in the bottom left of each panel.}
\label{fig:all_res}
\end{figure*}

\noindent Adopting the rotation curve from \cite{meidt13},
we generate a 2D model of pure circular motion using the \verb"GIPSY" task \verb"VELFI". This model
is subtracted from the observed velocity fields to obtain residual fields for PAWS at 1", shown in Fig~\ref{fig:paws_res}, and for the 30m, HERACLES and THINGS velocity fields, shown in Fig.~\ref{fig:all_res}. 
In the case of pure circular motion the residuals would be zero everywhere.  But here,  
residual velocity fields from each of the different surveys exhibit clear signatures of significant
non-circular
motions, with typical values between -30 and 30 km\,s$^{-1}$ and extrema reaching values above 90
km\,s$^{-1}$ (corresponding to the nucleus).  
In presence of density-wave structures, the non-circular motions introduce a particular
morphological pattern in the residual velocity field, as realized by \cite{canzian93}. 
In the case of a m=2 perturbation to the gravitational potential (introduced by a two-armed stellar
spiral or a stellar bar), the residual velocity field exhibits an m=1 pattern (i.e. an
approaching-receding dipole) inside corotation, and this changes to an m=3 morphology outside
corotation. This morphology shift is due to the change in sign of the gas streaming motions beyond
the corotation circle, affecting only their radial components, and is expected to appear at the
corotation only if the spiral structure is density-wave in nature with a constant pattern speed.\\

\noindent Although the pattern predicted by \cite{canzian93} can be difficult to distinguish at lower
spatial resolution, the residual velocity fields from the PAWS data at 1" and 3"  resolution (top of Fig. \ref{fig:paws_res}) show the signature very clearly, over several radial zones.  In the central
region ($R<35"$) the residual velocity field presents
a clear m=1 pattern consistent with motions driven by the m=2 stellar nuclear bar.  Just outside the
molecular ring at R=23'' and up until $R\approx$55", we see another approaching-receding dipole, now
introduced by inflow motions driven by the two-armed spiral in this region (especially clear at the
location of the southern spiral arm).  This is complimented by transition to an m=3 pattern beyond
$R<$55", although between this radius and  $R\lesssim70"$ the morphology
becomes more complex.  
In the outermost region ($R\gtrsim70"$), where the density-wave spiral transitions to material
spiral arms (\citealt{meidt13}), the PAWS FoV exhibits only a dipole.\\  

\clearpage
\newpage

\clearpage
\newpage

\clearpage
\newpage

\subsection{Harmonic decomposition of the non-circular velocity component}\label{sec:harm_dec}
 
\noindent In the previous section we identified several kinematic features not associated with pure
circular motion.\\ 

\noindent Here we use a powerful technique first introduced by \cite{schoenmakers97} to
describe and quantify non-circular motions, namely by expanding the peculiar component of the
line-of-sight velocity $V_{pec}$ as the harmonic series

\begin{equation}\label{eq:harm_exp} 
  V_{pec} = \sum_{j=1}^{N}[c_{j}\cos(j\theta)+s_{j}\sin(j\theta)]\sin(i),
\end{equation}

\noindent Here $N$ is the number of harmonics considered and $c_{j}$ and $s_{j}$ are coefficients
that describe the radial and azimuthal components of the non-circular motion, which can be
interpreted in terms of perturbations to the gravitational potential.  \cite{canzian93} showed
that
a potential perturbation of $m$ order introduces $j=m-1$ and $j=m+1$ patterns
in the residual velocity field, each on either side of the pattern's corotation radius (see the upcoming
section).\\  

\noindent We quantify the magnitude, or power, of each individual order of the harmonic decomposition $j$ as
the quadratically-added amplitude (e.g. \citealt{trachternach08}):

\begin{equation}\label{eq:Ai}
 A_{j} = \sqrt{c_{j}^{2}+s_{j}^{2}}.  
\end{equation}

\noindent and write the total power of all non-circular harmonic components as

\begin{equation}\label{eq:Ar}
 A_{r} = \sqrt{\Sigma_{j=1}^{N}[c_{j}^{2}+s_{j}^{2}]},
\end{equation}

\noindent to get a sense of the total magnitude of non-circular streaming motions.  In the next
section we
inspect radial trends in $A_{j}$ and $A_{r}$ for coincidence with morphological features in M51.
Later in Section~\ref{sec:vsp_omegap} we use our measurements of $A_j$ to calculate the magnitude of
the streaming motions associated with perturbations with $m$-fold symmetry.\\ 

\subsubsection{Application to residual velocity fields}
\noindent We perform the harmonic decomposition of the residual velocity field from PAWS at
1", PAWS 3", THINGS, HERACLES, and 30m velocity field up to order
$j=6$ using a modified version of the code first presented in \cite{fathi05}. The inclination and
PA of the best fitting ellipses are fixed to the values derived in
Section \ref{sec:kinpar} ($i=22^{\circ}$, $PA=173^{\circ}$) and the ring width is set to one
beam. Fig~\ref{fig:paws_res} and Fig~\ref{fig:all_res} shows the residual velocity fields
reconstructed from the harmonic decomposition (middle row). Since the difference between residual
velocity fields and the reconstructed fields is generally close to zero everywhere
(Fig~\ref{fig:paws_res} and Fig~\ref{fig:all_res}, bottom row) we are confident that the harmonic
decomposition using only 6 terms is quite accurate.\\

\noindent In
Fig.~\ref{fig:ampl_paws}, \ref{fig:ampl_things} and \ref{fig:ampl_hera} we plot the power in the
single
harmonic components, and their total, as a function of radius (bottom plot, top left and top right columns), the
median of these across the environments defined in \cite{meidt13} (e.g. nuclear bar, molecular ring, density-wave spiral arm and material arm regions; top plot, bottom left
and bottom right columns) and the median across the FoV (top plot). The error bars shown there
are obtained through a bootstrap technique. We generate 100 residual velocity fields, and 100
harmonic decompositions, for a range of PA and $i$ (set to their respective error bars).  We take
the results determined at our optimal $PA=173^{\circ}$ and $i=22^{\circ}$ as our final estimate and
define the error on that estimate as the median absolute deviation of the
bootstrapped amplitudes.\\ 

\noindent To discriminate between real trends and noisy peaks in the harmonic decompositions, we set a
confidence level at $2\times$ the channel width of the survey (i.e. 10 km s$^{-1}$ or in the case
of HERACLES 5.2 km s$^{-1}$). The (azimuthally averaged) harmonic components are highly reliable
when they are above this threshold.\\ 
 
\subsection{Global Trends}
\noindent As expected, surveys with high spatial resolution reveal larger streaming motions than those with
lower resolution.  
In PAWS 1" and PAWS 3" data the global amplitude of the non-circular components is $\langle
A_{r}\rangle$$\sim45$
km\,s$^{-1}$, whereas $\langle
A_{r}\rangle\sim20-35$
km\,s$^{-1}$ ~for the low resolution surveys, even when restricting the FoV to
the
PAWS FoV.  This difference stems from the fact that contributions from motions induced by the
nuclear bar and spiral
arms are not well resolved in these other surveys.\\

\noindent However all surveys, independent of resolution, very clearly show the signature of a dominant
two-armed pattern.  As predicted by \cite{canzian93} the expected $j=1$ and $j=3$ modes induced
by the
bar and two-armed spiral in M51 are apparent in all surveys: 
$j=1$ is the dominant mode of the residuals ($\langle A_{1}\rangle\approx30$ km s$^{-1}$ for PAWS
and $\langle A_{1}\rangle\approx10-20$ km s$^{-1}$ for the low resolution
surveys, approaching the total power within maps restricted to the PAWS FoV), followed by the $j=3$ mode ($\langle A_{3}\rangle\approx20$ km
s$^{-1}$ for PAWS
and $\langle A_{3}\rangle\approx10-15$ km s$^{-1}$ for the low resolution surveys).  However in all
cases, the $j=2$
mode has a value quite close to the $j=3$ ($\langle A_{2}\rangle\approx12$ km s$^{-1}$ for PAWS
and HERACLES maps and $\langle A_{2}\rangle\approx10$ km s$^{-1}$ for THINGS and PAWS single dish). 
A non-negligible $j=2$ velocity term would indicate a possible $m=1$ or $m=3$ perturbation to the
galactic potential.  However this is difficult to confirm from global measurements since, on
average, perturbations of order $j>3$ all have amplitudes $<10$ km s$^{-1}$. Given that individual
components may or may not extend as far as the
dominant two-armed spiral (that spans the entire field of view), below we
explore the evidence for $m=1$ and $m=3$ modes by analyzing radial trends.\\
  
\subsection{Radial Trends}
\noindent The high resolution of the PAWS data (at either 1" or 3") provides the most 
accurate depiction of the radial variation in the different harmonic components (at least for radii
$R<85"$).  We therefore focus on these data in this Section, but note similar trends when present in
the lower resolution survey data.  
\subsubsection{Odd velocity modes: the bar and two-armed spiral arms}
The innermost region of M51 ($R<23.5"$) is dominated by the peculiar motions driven by the
nuclear bar, which introduces a 
$j=1$ mode between 2 to 3 times stronger than the other modes in this zone
($\langle A_{r}^{(R<23.5")}\rangle\sim\langle A_{1}^{(R<23.5")}\rangle\sim35$ km\,s$^{-1}$).  Just
outside the bar, in the zone of the molecular ring ($23.5"<R<35"$), the peculiar motions are
reduced, reaching their lowest values across the FoV ($
A_{r}^{(23.5"<R<35")}\sim20$ km\,s$^{-1}$ and $A_{1}^{(23.5"<R<35")}\sim10$
km\,s$^{-1}$). 
However, near $R$=35'' the $j=1$ term begins to increase again ($\langle
A_{3}^{(23.5"<R<35")}\rangle\sim40$ km\,s$^{-1}$). After $R\sim$60''  the
power in the $j$=3 mode also once again increases, to a level comparable to that in the $j$=1 mode.\\  

\noindent Here the harmonic expansion confirms the visual impression from the residual velocity field
morphology analysis: inside the torque-based estimate of the first spiral arm corotation radius
($R_{CR}=55"$, \citealt{meidt13}) the residual velocity field appears dominated by a dipole pattern
($\langle
A_{1}^{(35"<R<55")}\rangle\sim40$ km\,s$^{-1}$ and $\langle A_{3}^{(35"<R<55")}\rangle\sim15$
km\,s$^{-1}$), while
beyond the $j=3$ term is stronger ($\langle A_{1}^{(55"<R<85")}\rangle\sim10$ km\,s$^{-1}$ and
$\langle A_{3}^{(55"<R<85")}\rangle\sim50$ km\,s$^{-1}$) and then reduces to $\sim10$
km\,s$^{-1}$ in the region $(65"<R<80")$.  
The switch in dominance from $j$=1 to $j$=3 in the PAWS 1'' and 3'' fields moreover occurs across a
zone that is consistent with the expected location of the corotation radius determined from
gravitational torques.\\

\noindent The existence of a  transition between a $j=1$ to a $j=3$ term is also clear at lower resolution, but
now the transition occurs slightly further out at $R$$\sim$70" in HERACLES and 30m data. This
displacement in the position of the transition with respect to the transitions in PAWS at 1" and PAWS at 3" could be caused by
beam smearing that extends the transition radius over a wider region. However this switch in
dominance in not well defined in THINGS 6".\\

%\newpage

\begin{figure*}[!h]
\begin{center}
\includegraphics[width=0.5\textwidth]{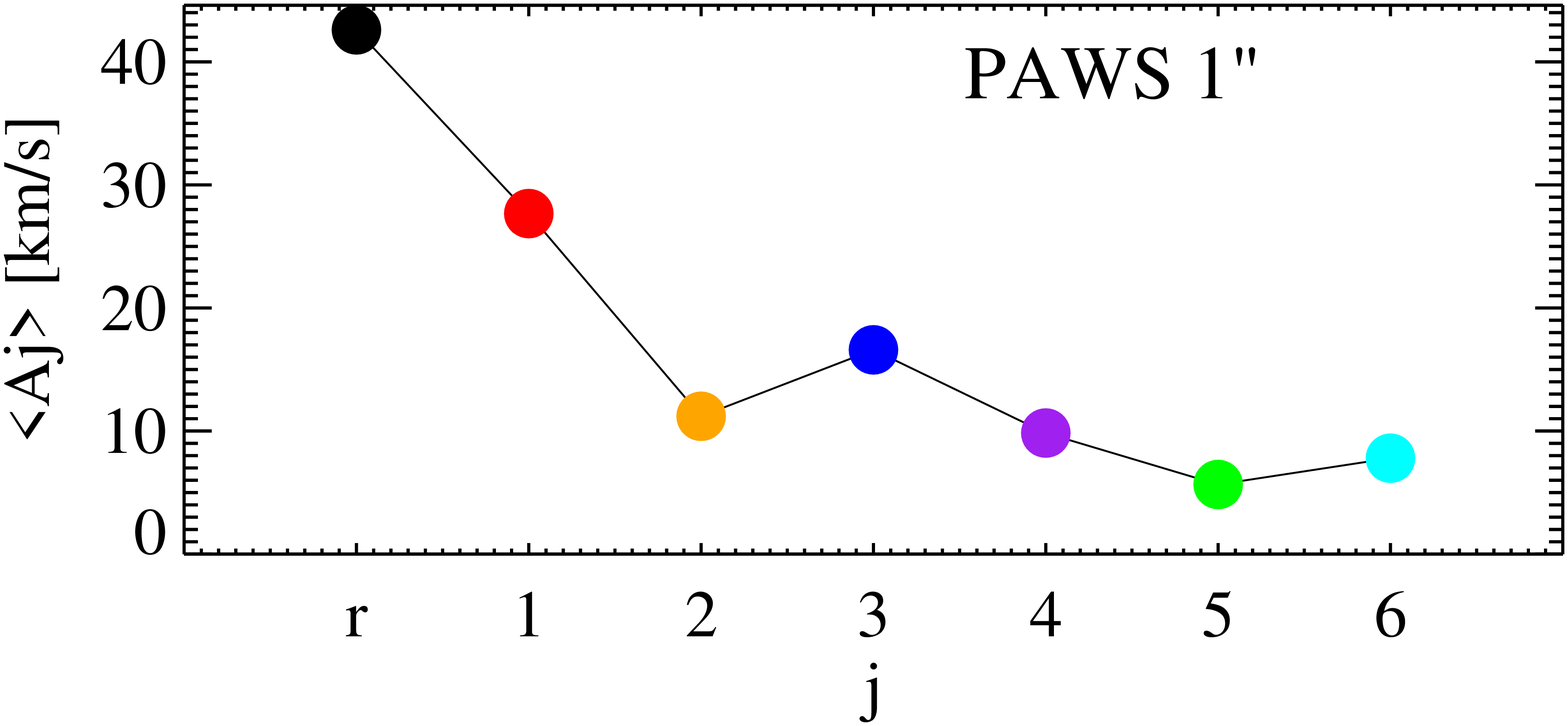}
\includegraphics[width=1\textwidth]{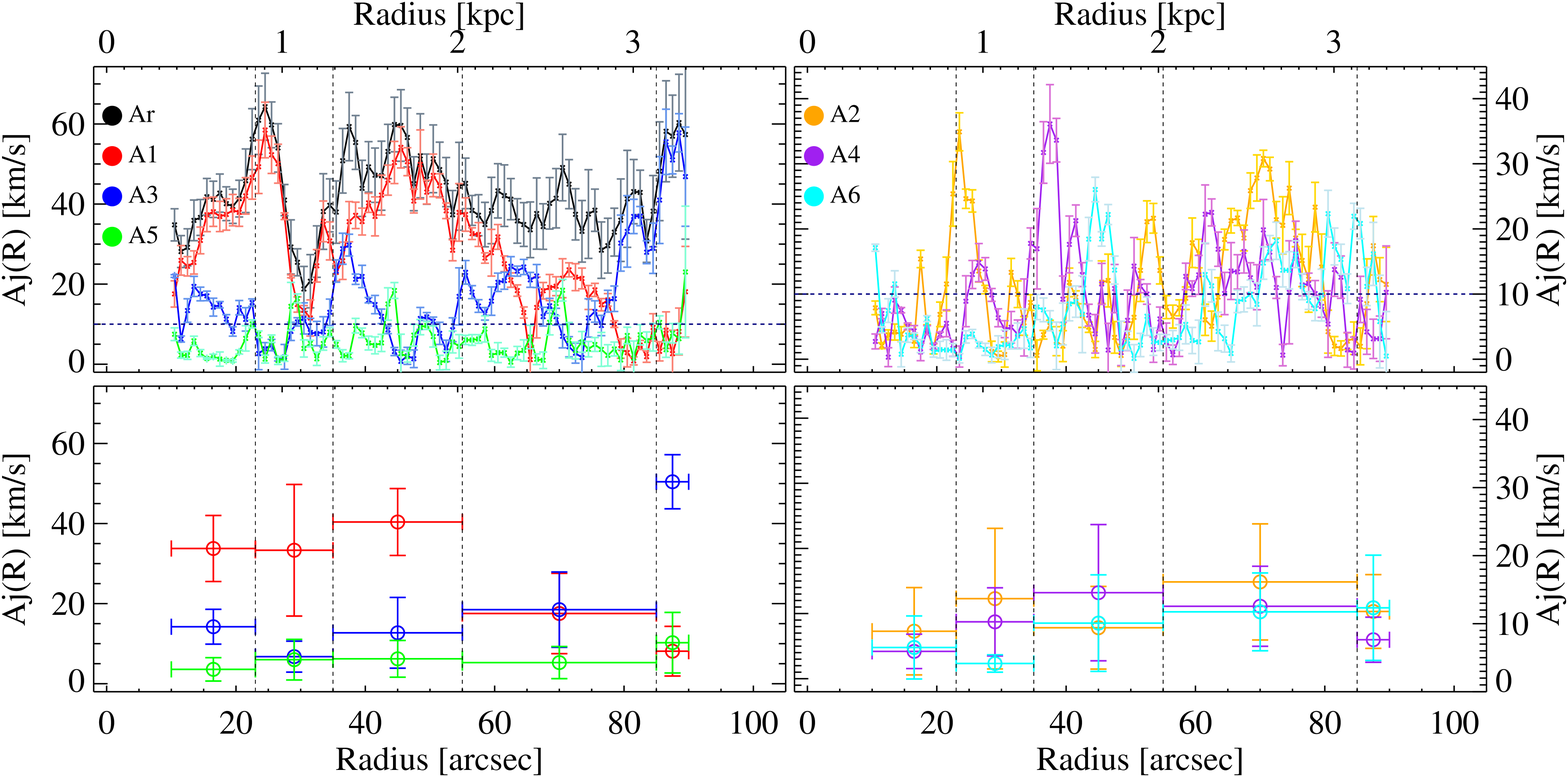}
\end{center}
\caption{\footnotesize \emph{Top plot:} Radially averaged mean of the harmonic component amplitudes
$A_{j}$
from PAWS 1" residual velocity field. 
\emph{Bottom plot:} Non-circular motion amplitudes from harmonic decomposition: radial
trend of the odd components and the total power $A_{r}(R)$ (top left) and even components
(top right). The horizontal blue dashed straight line indicates twice the channel width of the datacube, i.e. $2\times5$ km\,s$^{-1}$ = 10 km\,s$^{-1}$. In the bottom row
the mean behavior of the odd (left) and even (middle) components in the different M51 environments
 as defined in \cite{meidt13} (dashed vertical lines; see the text for details) are indicated together with the standard deviations of the values. Horizontal error bars represents the widths of the environments.}
\label{fig:ampl_paws}
\end{figure*}

%\clearpage
%\newpage

\subsubsection{Even velocity modes: an additional three-armed spiral structure}
\noindent The higher resolution maps also provide valuable information about other, weaker modes
that appear over a more limited radial range than those associated with the dominant two-armed
pattern. Compared to lower spatial resolution data, we can sample this type of mode in PAWS data at
1'' and 3'' with many more resolution elements.\\

\noindent Fig.~\ref{fig:ampl_paws} shows that there is non-negligible power in several of the even harmonic
components, over almost the entire PAWS FoV.  The $j$=2 exhibits a strong peak of $\sim35$ km
s$^{-1}$ at
$R\approx23"$. 
Between $25"\lesssim
R\lesssim40"$ the $j$=2 term weakens and the power in the $j$=4 term increases, peaking well above
our
confidence level ($\sim35$ km s$^{-1}$ at $R\approx37"$). This switch in dominance between $j$=2 and
$j$=4 term is most clear in the PAWS 1" velocity field.\\   

\noindent Since a perturbation of $m$ order introduces $j=m-1$ and
$j=m+1$ terms in the residual velocity field, non-negligible values of $j$=2 and $j$=4 constitute the first kinematic evidence of
an $m=3$ wave
within $R\sim45"$ (i.e. $R\sim1.7$ kpc) in the disk of M51a.  According to the transition between
these two
components, we estimate that the corotation radius of this mode occurs at $R=(30\pm3)"$ (i.e.
$R=1.1\pm0.1$ kpc\footnote{The corotation radius of the m=3 mode has been fixed to
the center of the region where $j$=2 and $j$=4 overlap. The uncertainty is given by the width of
this zone.}).\\ 

\noindent The PAWS data at 3" show a similar pattern, including a switch in dominance between $j$=2 and $j$=4
term
at a similar radial distance as in PAWS 1".   But given the lower resolution, the detection of the
$j$=4 in the region
between $45"\lesssim
R\lesssim50"$ occurs over only 5
data points, and the signature is also weaker (the maximum is $\sim25$ km s$^{-1}$).  Moving
to resolution lower than 3", the behaviors of $j$=2 and $j$=4 terms are gradually smeared out and
the switch in dominance between the two modes is no longer obvious.\\

\noindent An $m$=5 potential perturbation could also be responsible for the $j$=4 term.  But, in this case we
would expect a more substantial $j$=6 term at larger radii than is measured; only few data
points of the $j$=6 term have values above our confidence level.  We therefore conclude that this scenario is
improbable, or is difficult to detect with the present (spatial and spectral) resolution.\\  

\noindent Likewise, since the $j$=2 component, which becomes dominant again outside $R$$\sim$2 kpc, is never
accompanied by another transition to a $j$=4 mode with significant power at larger radii, we argue
that this must
describe a genuine lopsidedness arising with an $m$=1 perturbation.\\  

\subsubsection{Outer arms}
\noindent In the region corresponding to the material arms the PAWS FoV has few data points and the
decomposition becomes less accurate. Here it is useful to consider the results from the other lower
resolution surveys\footnote{The resolution of the $30m$ dataset is too coarse for this kind of analysis and so we do not consider it here.}. 
The total power of the non-circular components $A_{r}(R)$ increases almost monotonically in all
harmonic expansions, from 10-20 km s$^{-1}$ in the innermost region to $\sim30$ km s$^{-1}$ at
140".  In the HERACLES 13.5" map the $j=3$ remains dominant across the whole FoV, with
$\langle A_{3}\rangle\sim20-30$ km s$^{-1}$.\\

\begin{figure}[!h]
\begin{center}
\includegraphics[width=0.75\textwidth]{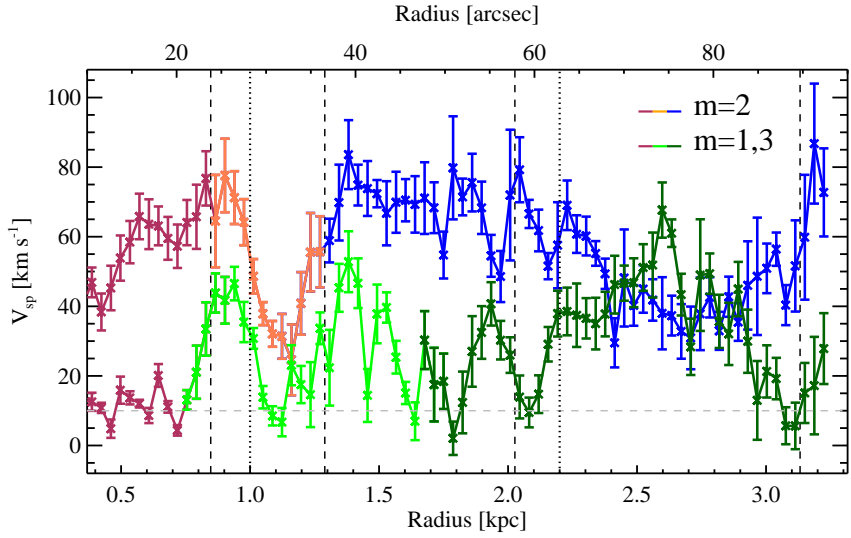}
\end{center}
\caption{\footnotesize Amplitude of the spiral perturbation from PAWS 1" from the main two-fold spiral arms, three-fold structure and $m=1$ mode. Dark red indicates the radial extent of the bar, orange the molecular ring region, and blue the density-wave spiral arm region as identified from the present-day torque analysis by \cite{meidt13}. Light green indicates the region of a possible $m=3$ influence, and dark green the non-circular motion induced by the m=1 perturbation. Dotted vertical lines represent the region where \cite{henry03} observed a strong signature for a deviation from a pure $m=2$ symmetry. Vertical dashed lines indicate the M51 environments as defined in \cite{meidt13} (see the text for details).}
\label{fig:vsp_paws}
\end{figure}

\clearpage
\newpage

\subsection{The magnitude of streaming motions}\label{sec:vsp_omegap_theo}

\noindent In the previous two sections we used measurements of the power in individual components of the
harmonic expansion of  the residual line-of-sight velocities observed in M51 to characterize the
non-circular motions driven by non-axisymmetric structures.  
In this section we will give these a physical interpretation, which we will then use to understand
the nature of M51's patterns.\\  
 
\noindent Similarly to \cite{wong04}, we express the peculiar velocity component $V_{pec}$ in
Eq.~\ref{eq:vpec2} in terms of the velocities driven in response to a spiral perturbation to the
gravitational potential with $m-$fold symmetry, following \cite{canzian97}:
\begin{equation}\label{eq:vpec2}
  V_{pec}=V_{sp}\left[\frac{\kappa}{2\Omega}\cos(\theta+\chi)\sin
m(\theta-\theta_{sp})+\nu\sin(\theta+\chi)\cos m(\theta-\theta_{sp})\right]\sin i,
\end{equation}
Here, $v_{sp}$ is the velocity amplitude that depends on the magnitude of the spiral perturbation,
$\theta_{sp}$ is the spiral phase, $\chi$ the spiral arm pitch angle (the angle between the tangent
to the arm and a circle with constant radius; by definition $0^{\circ}<\chi<90^{\circ}$) and
assuming S-spiral symmetry and trailing spiral arms in the case of M51\footnote{An S-spiral has a
shape like the letter ``S''. This convention refers to the two projections of a (trailing-arm)
spiral on the plane of the sky. For details see \cite{canzian97}}.   The angular frequency
$\Omega\equiv(V_{c}/R)^{-1}$,
with $R$ the galactic radius in kpc, the pattern speed of the spiral arms is $\Omega_{p}$ and the
dimensionless frequency $\nu$ and epicyclic frequency $\kappa$ are
defined as  

\begin{equation}\label{eq:nu_kappa}
 \nu\equiv\frac{m(\Omega_{p}-\Omega)}{\kappa},
\quad\kappa^{2}\equiv4\Omega^{2}+R\frac{d\Omega^{2}}{dR}.
\end{equation}

\noindent As shown by \cite{wong04}, in the case of a single perturbation with mode $m$, the
harmonic decomposition of the
peculiar velocities in Eq.~\ref{eq:vpec2} yield harmonic coefficients of the form:

\begin{eqnarray}\label{eq:vpec_hd}
 c_{m\pm1}=\frac{V_{sp}}{2}\left(\frac{\kappa}{2\Omega}\pm\nu\right)\sin(m\theta_{sp}\pm\chi),\\
 s_{m\pm1}=\frac{V_{sp}}{2}\left(\frac{\kappa}{2\Omega}\pm\nu\right)\cos(m\theta_{sp}\pm\chi).
\end{eqnarray}

\noindent In the general case of more than one mode $m$, each with its own unique pattern speed
$\Omega_{p,m}$, $\chi_m$ and $\theta_{sp,m}$, and which each drives its own streaming motions with amplitude
$V_{sp,m}$, we can express the amplitudes of any set of  harmonic components as

\begin{eqnarray}\label{eq:vpec_A}
 A_{m\pm1} = \sqrt{c_{m\pm1}^{2}+s_{m\pm1}^{2}} =
\frac{V_{sp,m}}{2}\left(\frac{\kappa}{2\Omega}\pm\nu_m\right).
\end{eqnarray}

\noindent Combining $A_{m-1}$ and $A_{m+1}$ with the definition of the dimensionless frequency $\nu_m$
in
Eq.~\ref{eq:nu_kappa} we can obtain the following simple parametrization of the amplitude of
velocity perturbation: 
\begin{equation}\label{eq:vsp}
 V_{sp,m}=\frac{2\Omega}{\kappa}(A_{m-1}+A_{m+1}).\\
\end{equation}

\noindent The linear combination of $j$=1 and $j$=3 amplitudes, for instance, provides a measure of the streaming
motions driven by an $m$=2 spiral perturbation. In this way, in the presence of more than one mode we can isolate the contributions of individual modes to the total observed non-circular motions.  This method for measuring streaming motions also does not need to assume a specific spiral arm pitch angle (observed to vary in M51, e.g. \citealt{schinnerer13}) to perform the decomposition, as required by the technique employed by \cite{meidt13}.\\

\noindent Similarly, the spiral arm pattern speed $\Omega_{p}$ can be expressed as

\begin{equation}\label{eq:omegap}
 \Omega_{p,m}=\frac{\kappa}{m}\left(\frac{A_{m+1}-A_{m-1}}{v_{sp,m}}\right)+\Omega. 
\end{equation}

\noindent Note that when $A_{m+1}=A_{m-1}$, $\Omega_{p}=\Omega$.  This is a recasting of the
prediction by \cite{canzian93} that \emph{corotation} radius (where $\Omega_{p}=\Omega$) is
crossed
when the $m-1$ switches to an $m+1$ term.   
However, we emphasize that the pattern speed is likely impossible to estimate reliably in this way,
since it depends on $\kappa^2$; $\kappa$ itself can be difficult to accurately constrain with
observation and is susceptible to uncertainty as it depends on the derivative of $\Omega$ (see
Eq.~\ref{eq:nu_kappa}). 
For a recent estimation of the radial variation of the spiral arm pattern speed in M51a
through the more reliable and model-independent radial Tremaine-Weinberg (TWR) method, we refer the
reader to \cite{meidt08}.\\

\subsubsection{Streaming motions in M51}\label{sec:vsp_omegap}
\noindent  In this section we use the results of the harmonic decomposition and our model of M51's
rotation curve to estimate  the magnitude of streaming motions (Eq~\ref{eq:vsp}) driven in
response to the bar, dominant two-armed spiral, the three-armed spiral pattern and/or $m=1$ mode.\\  

\noindent We start considering solely the $m=2$ perturbation of the galactic potential. 
In this case, the quantity of interest is obtainable from the $A_{1}$ and $A_{3}$ as:

\begin{equation}\label{eq:vsp_m2}
 V_{sp,m=2}=\frac{2\Omega}{\kappa}(A_{1}+A_{3}),
\end{equation}

\noindent where $\Omega=V_{c}/R$ and $\kappa$ is given by Eq~\ref{eq:nu_kappa}.\\

\noindent  Fig.~\ref{fig:vsp_paws} and Fig.~\ref{fig:vsp_all} show the amplitude of velocity of the
spiral arm perturbation as
derived from Eq.~\ref{eq:vsp_m2} using the harmonic amplitudes from PAWS 1" and lower resolution data residual velocity fields, respectively, as analyzed in Section~\ref{sec:harm_dec}.  In the nuclear bar region ($R<23"$) streaming motions are
$\langle V_{sp,m=2}(R<23")\rangle\approx60$ km s$^{-1}$, in the PAWS 1'' data set. Further
the streaming motions reach the highest values with a median of $\langle
V_{sp,m=2}(35<R<60")\rangle\approx70$
km
s$^{-1}$ in PAWS 1" than it decreases again to values around $V_{sp,m=2}(60<R<85")\rangle\approx50$
km s$^{-1}$.  However in the lower resolution
surveys (i.e. THINGS 6",
HERACLES 13.5" and PAWS single dish 22.5"), $\langle V_{sp,m=2}\rangle$ is always below $\sim50$
km
s$^{-1}$ and reaches a value comparable to that recorded in PAWS only in the region of the material
arms ($R>85"$). 
This behavior could be due to beam
smearing that reduces the observed peak in streaming motions.  As discussed in
Section~\ref{sec:disc_dataset}, in the case of HI, this could be also due to an intrinsically
different response to the spiral perturbation of the potential.   
In all cases, the spiral perturbation velocity drops in the molecular ring region reaching the
minimum of $V_{sp,m=2}\approx25$ km s$^{-1}$ for PAWS 1", as expected from an analysis of 
gravitational torques (\citealt{meidt13}).\\

\noindent In Fig.~\ref{fig:vsp_paws} and Fig.~\ref{fig:vsp_all} we plot also the radial profile of streaming motions that
corresponds to
the $m$=3 and $m=1$ perturbations, calculated according to:
\begin{equation}\label{eq:vsp_m3}
 V_{sp,m=1,3}=\frac{2\Omega}{\kappa}(A_{2}+A_{4}),
\end{equation}

\noindent As described in the previous section, we expect these motions to be related to the $m=3$ wave
between $20"<R<45"$ (i.e.,
$0.8$ kpc $<R<1.7$ kpc), where we observe a peak in the $j$=2 term switching to a peak in a $j=4$
term in the residual velocity field.  The start of the $m$=3 mode is taken as the location where the
$j$=2 term increases above our 10 km s$^{-1}$ confidence threshold, while the end of the $m$=3 mode
is set by the decrease in the power of the $j$=4 term.  This zone is consistent with the radial
range over which the larger deviation from a pure $m=2$ mode
was identified ($1$ kpc $<R<2.2$ kpc, \citealt{henry03}).   Across this zone, the $m$=3 mode drives
streaming motions of $\langle V_{sp,m=3}\rangle\approx25-30$ km s$^{-1}$ on average and reaches a minimum below the confidence limit of 10 km s$^{-1}$ in the ring region. 
(Note that there is little to no power in the zone of the
bar where $\langle V_{sp,m=1,3}\rangle\approx12$ km s$^{-1}$, only slightly above our confidence limit). 
At larger radii, the streaming motions arise from a lopsided ($m$=1) mode (only
$j=2$ appears in the harmonic expansion, i.e. $A_{4}\sim$0), with a magnitude of $\langle
V_{sp,m=1}\rangle\approx32$ km s$^{-1}$.\\

\newpage

\section{Discussion: an $m$=3 potential perturbation in M51}\label{sec:disc_m3}
\noindent In the previous sections we presented kinematic evidence for the existence of an $m$=3
mode, which supplies confirmation of an $m$=3 perturbation to M51s gravitational potential first
investigated by \cite{elmegreen92}.  This mode is spatially coincident with
the inner part of the dominant two-armed spiral.  Presumably, the interference of an $m=3$ wave with
the $m=2$ wave enhances the asymmetry in the velocity field (i.e. increasing the deviation in 
iso-velocity contours from pure circular motion.)  This would seem to support the interpretation of
\cite{meidt08}, who consider the likelihood that their inner TWR pattern speed estimate calculated
using
CO(1-0) as a kinematic tracer reflects a combination of the speed of the $m$=3 mode with that of the
dominant two-armed spiral.\\

\noindent This conclusion moreover supports the finding of \cite{henry03}, who reconsidered the evidence
for an $m$=3  perturbation in the old stellar light distribution first studied by \cite{rix93}.
They claim that the magnitude of the $m=3$
component in K-band is sufficient to account for the offset between the mirror of one of the two
main spiral arms and its counterpart.   They also observe patches of molecular gas and star
formation in the inter-arm at the location of one of the three arm segments imaged in the K-band.\\

\noindent In the next section we consider the origin of this $m$=3 mode and its density-wave nature, taking
into account our analysis of the gas response.\\

\begin{figure}
\begin{center}
\includegraphics[width=1\textwidth]{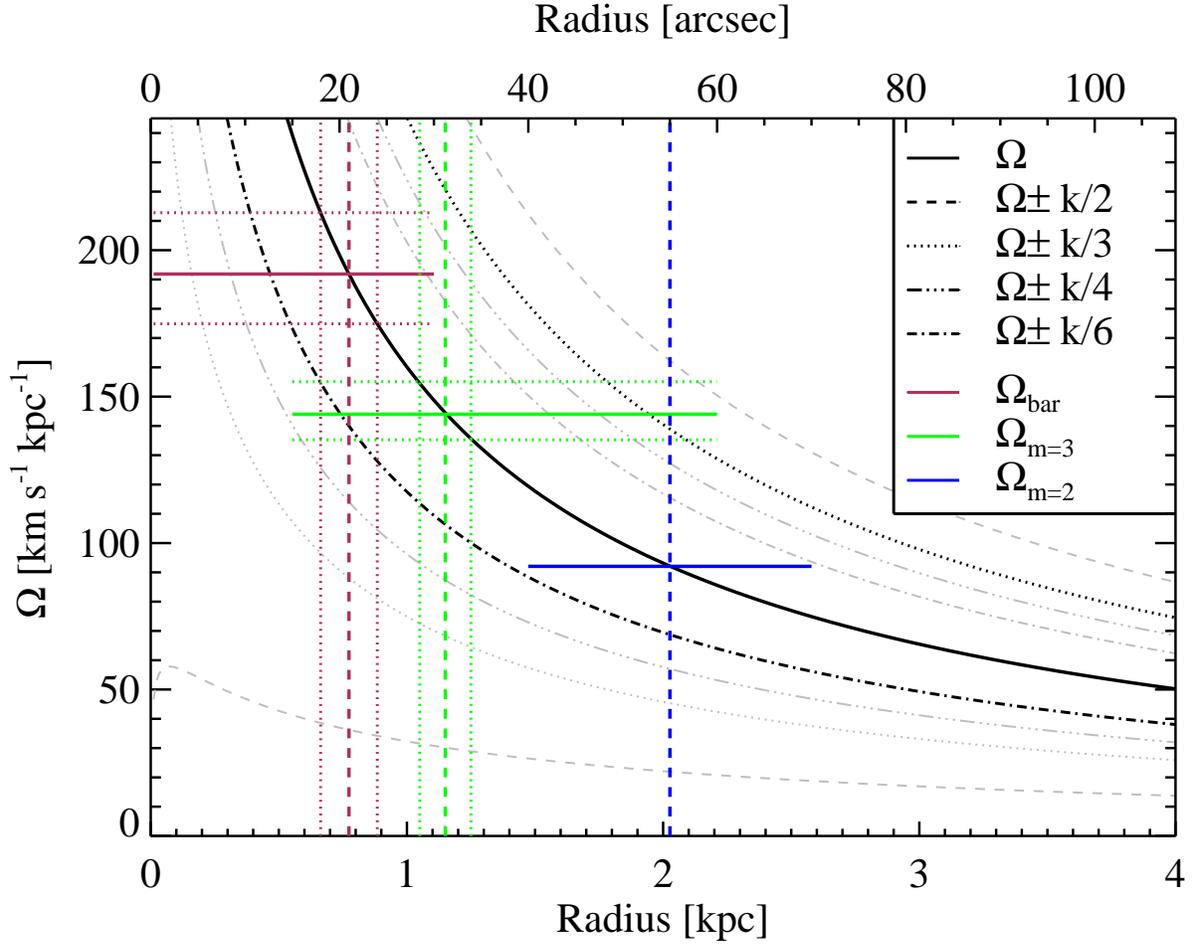}
\end{center}
\caption{\footnotesize Angular frequency curves derived from the gas-based rotation curve of M51:
$\Omega$ (black solid), $\Omega\pm\kappa/2$ (dashed), $\Omega\pm\kappa/3$ (dotted),
$\Omega\pm\kappa/4$ (dashed dotted-dotted-dotted)
$\Omega\pm\kappa/6$ (dashed-dotted). Angular frequency curves discussed specifically in the text ($\Omega-\kappa/6$, $\Omega+\kappa/3$) are highlighted in black. Pattern speed estimates for the nuclear bar, spiral
arms, and m=3 density-wave in M51 are shown in dark red, blue and green, respectively,
together with their associated corotation radii and uncertainties (when available).}
\label{fig:omega_m3}
\end{figure}

\subsection{Origin, role and nature of the $m=3$ mode}

\begin{figure}
\begin{center}
\includegraphics[width=0.75\textwidth]{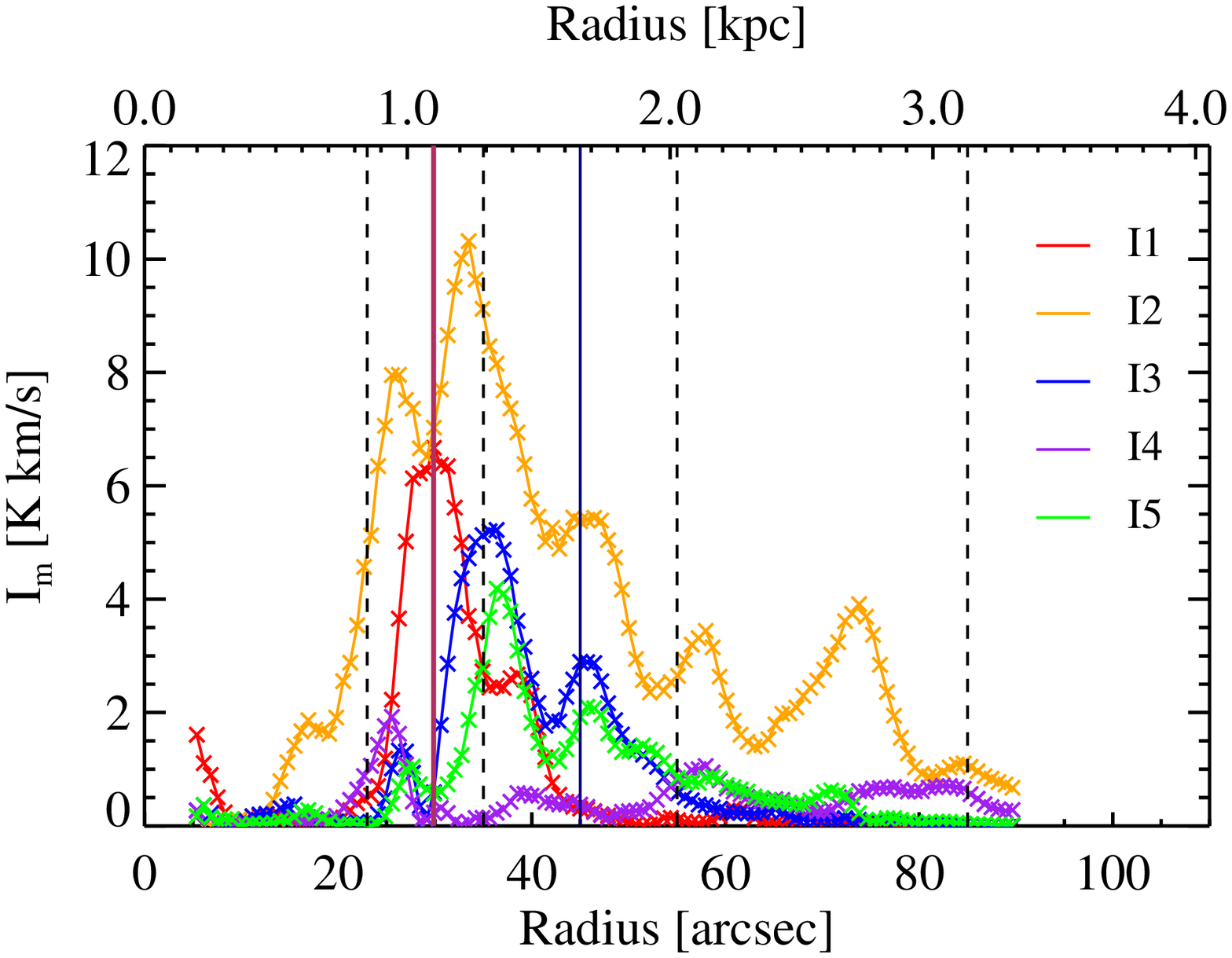}
\end{center}
\caption{\footnotesize Fourier decomposition of the surface brightness of the PAWS 3" zeroth
moment map shown as the power in the Fourier component in K\,km\,s$^{-1}$. The vertical blue line indicates the boundary between $m=3$ and $m=1$ dominance estimated in Section~\ref{sec:vsp_omegap}. The red vertical line represents the $m=3$ corotation at $R\sim1.1$ kpc. Dashed vertical lines indicates M51's environments as defined in \cite{meidt13}.}
\label{fig:hd_mom0}
\end{figure}

\noindent The PAWS 1" residual velocity
field shows a clear kinematic signature of an $m=3$ mode in the
central region of M51a. According to Fig.~\ref{fig:ampl_paws} we place its corotation radius at
$R_{CR,m=3}\approx(30\pm3)"$ (i.e. $R_{CR,m=3}\approx1.1\pm0.1$ kpc). Together with the angular
frequency derived by \cite{meidt13}, we can define the pattern speed of $\Omega_{m=3}\approx140\pm9$ km s$^{-1}$ kpc$^{-1}$.\\

\noindent Fig.~\ref{fig:omega_m3} shows that the $m=3$ wave appears to be associated with several interesting resonance overlaps, giving us a picture of very specific interaction between waves. The corotation radius $R_{CR,b}\approx0.8\pm0.1$ kpc for the nuclear bar of M51 (\citealt{zhang12}) overlaps with the inner ultra harmonic resonance (UHR) of the $\Omega_{p,m=3}$ pattern speed (where $\Omega_{p,m=3}$=$\Omega-\kappa/6$). The $m=3$ mode itself extends out to R$\sim1.7$ kpc (according to where the amplitude of $j=4$ is above our confidence threshold), which is very close to the bar's outer Lindblad resonance (OLR), the outermost extent of its gravitational influence. This suggests that the bar is a possible driver of the $m=3$ mode. The $m=3$ mode also appears to be connected with the main spiral structure.   Indeed, the OLR of the $m$=3 (where $\Omega_{p,m=3} = \Omega + \kappa/3$) overlaps with the corotation radius of the main $m$=2 spiral pattern.\\ 

\noindent These resonance overlaps may be an instance of non-linear mode coupling. Fig.~\ref{fig:hd_mom0} presents the power in the Fourier decomposition\footnote{The Fourier
decomposition of the surface brightness is analogous to the harmonic decomposition of the
residual velocity fields performed in Section~\ref{sec:harm_dec}, but in this case the amplitudes of the
Fourier modes for $m=1-5$ are given by $I_{m}=\sqrt{s_{m}^{2}+c_{m}^{2}}$.} of the PAWS CO(1-0) surface brightness at 3", revealing power in both the $m=1$ and $m=5$ modes. This is in agreement with predictions by \cite{masset97} (and studied by \citealt{rautiainen99}) that coupling between
$m=2$ and $m=3$ modes should generate $m=1$ and $m=5$ beat modes.  The $m=1$ and $m=5$ modes are particularly strong and confined within the region of influence of the $m=3$ ($R<45"$). Moreover the $m=1$ mode is peaked exactly at the $m=3$ corotation of $R=1.1$ kpc.\\

\noindent This non-linear mode coupling can be interpreted as evidence that the particular $m$=3 structure we find provides the avenue to couple the bar, which we expect appears as a natural instability of the rotating stellar disk, with the dominant two-armed spiral that
extends out to larger radii, and which is presumably independently excited by the interaction with M51b. While bars and two-armed spirals are often suggested to naturally couple (in which case the bar is said to `drive' the spiral), in M51 this does not appear to be the case: Fig.~\ref{fig:omega_m3} shows no compelling direct link between the bar resonances (CR, OLR) and those of the $m$=2 spiral (ILR, UHR, CR).  The $m=3$, on the other hand, appear to supply a link between these two structures, presumably in order for energy and angular momentum to be continually transferred radially outward.\\

\noindent These evidences suggest that the $m=3$ mode as a density-wave nature. The transience or longevity of this feature, however, cannot be assessed with our observational data, which provides a snapshot of the current state of M51.  We note, though, that multiple spiral structures are generally associated with transient, quickly-evolving spiral arms (e.g. \citealt{toomre81}, \citealt{fuchs01}, \citealt{donghia13}).  
Since we would argue that the coincidence of a three-fold
potential perturbation with that of the main $m=2$ pattern definitively excludes a single mode in
M51 (like \citealt{lowe94}; \citealt{henry03}) our finding may therefore favor theories of multiple, quickly-evolving density wave spirals.\\

\noindent At larger radii, the residual velocity field harmonic decomposition indicates that the
$m=2$ wave may be spatially coincident with an $m=1$ perturbation to the potential.  This
perturbation is likely responsible for the lopsidedness in K-band images identified, e.g. by
\cite{rix93}. To reliably connect the origin of this feature to the interaction with M51b, new high
resolution data beyond the PAWS FoV are necessary.\\

\section{Discussion: The Dependence of Kinematic Parameters on Resolution and Gas Tracer}\label{sec:disc_dataset}
\noindent In the previous section we discussed
evidence for the existence of an $m=3$
wave in the radial range 0.8 kpc $<$$R$$<$1.7 kpc (i.e. $20<R<45"$)
in the center of M51a. The kinematic
signature of such a weak,
compact mode can be reliably identified only when analyzing 
the PAWS residual velocity field at a resolution of
1".  At lower spatial resolution
(even with equivalent spectral resolution), the presence 
of such a weak mode becomes less obvious 
(see Section~\ref{sec:harm_dec}). 
Given that the dominant molecular spiral arm 
width is around 400 pc (\citealt{schinnerer13}), it is
not surprising that
high resolution data are needed for an accurate 
kinematic characterization of the structures traced
by molecular gas.  
Other small scale kinematic features, such as 
the bright and high-velocity dispersion core of M51a
and
the spurs on the downstream side of the spiral arms, 
also only become visible in high
resolution velocity fields.\\

\noindent Perhaps more critical to the results of an in-depth kinematic analysis than resolution considerations is the nature and distribution of the kinematic tracer. Indeed, HI emission appears naturally more smooth at all spatial scales (\citealt{leroy13b}), which may make it less sensitive to small-scale potential perturbations than the highly clumpy medium traced by CO radiation.\\

\noindent For this reason, to correctly characterize spiral arm gas kinematics a gas phase tracer that is strongly affected by the mid-plane galactic potential and interferometric observations that are able to resolve them are preferred. In the following we illustrate how the nature of the tracer and the observing strategy for a given dataset impacts the interpretation of the kinematic properties measured for spiral galaxies like M51.\\

\subsection{CO versus HI kinematics}
\noindent Recent studies have shown that the 3-dimensional distributions of the atomic and molecular gas in M51 are not identical (e.g. \citealt{schinnerer13}, \citealt{pety13}). Therefore we expect to find differences in their kinematics as well. The CO line emission is closely associated with the spiral arms tracing the density enhancement in the old stellar population, while the emission from the atomic gas is fairly smooth and its brigthness distribution relative to the spiral arm suggests that it may be produced via the photodissociation of H$_{2}$. (e.g. \citealt{smith00}, \citealt{schinnerer13}, \citealt{louie13}). Moreover, the velocity dispersion observed in the CO-bright compact component emission is very different from the HI line emission, $\sim5$\,km\,s$^{-1}$ (\citealt{pety13}) versus $\sim15$\,km\,s$^{-1}$ (e.g. \citealt{tamburro09}, \citealt{caldu13}), respectively. According to \cite{koyama09} equation 2, this implies that the CO bright emission arises from a thinner disk than the HI radiation.\\

\noindent The different distribution of the molecular and atomic gas is also strongly reflected in their velocity fields from which all kinematic information is derived. As noted in Section 3, the PAWS CO velocity field tapered to 6" still shows several of prominent non-circular motion features that are clearly visible at 1" resolution while these features are basically absent in the THINGS HI cube at the same 6" resolution. (R2a) The first direct consequence is that rotation curves derived from CO and HI velocity fields are very different (Fig~\ref{fig:vsp_pt6as}, top). Rotation curves from PAWS CO datasets show strong bumps and wiggles at both 1" and 6" resolution. These features are mostly absent in the THINGS rotation curve, which is much smoother than the rotation curve obtained using the CO data.  In the latter, the presence of wiggles presumably reflects a contribution from azimuthal non-circular streaming motions in regions where the spiral arms dominate the tilted-ring fit compared to the relatively streaming-free inter-arm region.\\

\noindent For similar reasons, the residual velocity field from PAWS shows clear signatures of non-circular motion that are not present in the THINGS residual velocity field at the same resolution, pixel size and FoV (Fig.~\ref{fig:kin_comparison}, top left). Since those velocity fields are central to study spiral perturbations we illustrate their differences more quantitatively using pixel-by-pixel diagrams (Fig.~\ref{fig:kin_comparison}, top right). The pixel-by-pixel comparison reveals a large scatter between values measured in the two residual velocity fields . Such differences naturally influence the measurement of the velocity associated with the potential perturbation $V_{sp}$ (Fig~\ref{fig:vsp_pt6as}, bottom), which depends on the amplitude of (non-circular) harmonic components in the residual velocity field (see Eq.~\ref{eq:vsp}).  Whereas the magnitudes of the streaming motions derived using the PAWS 1" and 6" data are comparable, the value derived from the THINGS 6" data is on average $\
sim35$ km\,s$^{-1}$ lower than $V_{sp}$ obtained from PAWS 6" in the region between $R\sim60-80"$.\\ 

\noindent Our conclusion is that due to the different spatial distributions of the atomic and molecular gas (both in and above the disk plane), CO and HI emission trace the galactic potential differently.  Since the CO emission has a radial and vertical distribution that correlates very well with the location of the stellar spiral potential in M51, it is an optimal tracer for detailed kinematic characterization of the mid-plane potential.  Meanwhile, the atomic gas sits further away from the mid-plane and offset from the spiral arms so that it experiences a slightly different (and weaker) spiral perturbation.  As a result, CO is a better tracer of streaming motions, but HI yields better constraints on the bulk motion of the galaxy (i.e. the rotation curve and other global kinematic parameters).\\ 

\subsection{Hybrid versus single-dish data}
\noindent Interferometers filter out low spatial
frequencies, i.e., spatially extended emission. For this reason, the type of observational data that is used will affect the way a given gas phase observation traces motions driven in response to the gravitational potential.
Single dish observations are likely to be more sensitive to fluffy emission from a more vertically
extended component, as was recently discovered for the 30m and hybrid 30m+PdBI observations of M51
by \cite{pety13}.  As discussed at the end of the previous section, this may prevent
single-dish observations from revealing the same pattern of streaming motions that are evident even in the hybrid data after degrading its resolution.\\  

\noindent The middle row of Figure \ref{fig:kin_comparison} shows this in a little more detail, comparing the
PAWS and HERACLES residual velocity fields smoothed to the same 13.5"
resolution.\footnote{To put the two residual velocity field on the same resolution we smoothed PAWS
tapered at 6" to the HERACLES resolution of 13.5".} 
Even at 13.5", the PAWS residual velocity field
still exhibits the typical signatures of bar and spiral arm streaming motions.  But these departures
from circular motion are less clearly visible in the HERACLES residual velocity field.  The
pixel-by-pixel diagram confirms that the two maps are not the same, as large scatter is
present.\\

\noindent The line-width measured from HERACLES IRAM 30m observations is significantly larger than measured
from PAWS at 1".  Some part of this could be due to unresolved bulk motions.
\cite{caldu13} measured similar velocity dispersions for CO from HERACLES and HI from
THINGS observations in a sample of 12 galaxies, which would imply that the two phases have similar
vertical distributions. They find, for M51 in particular, $\sigma_{HI}\sim\sigma_{CO}\approx$ 15 km s$^{-1}$. 
This value is comparable to the velocity dispersion of the extended CO  component measured by \cite{pety13}
for M51, rather than the compact CO emission that dominates the PAWS second moment map (see \citealt{pety13}).  
This suggests that the single-dish data are dominated by the vertically
extended gas than the hybrid data, which mainly traces gas that is more
confined to the disk mid-plane, and thus more influenced by the gravitational potential.\\

\begin{figure*}[!h]
\begin{center}
\includegraphics[width=0.75\textwidth]{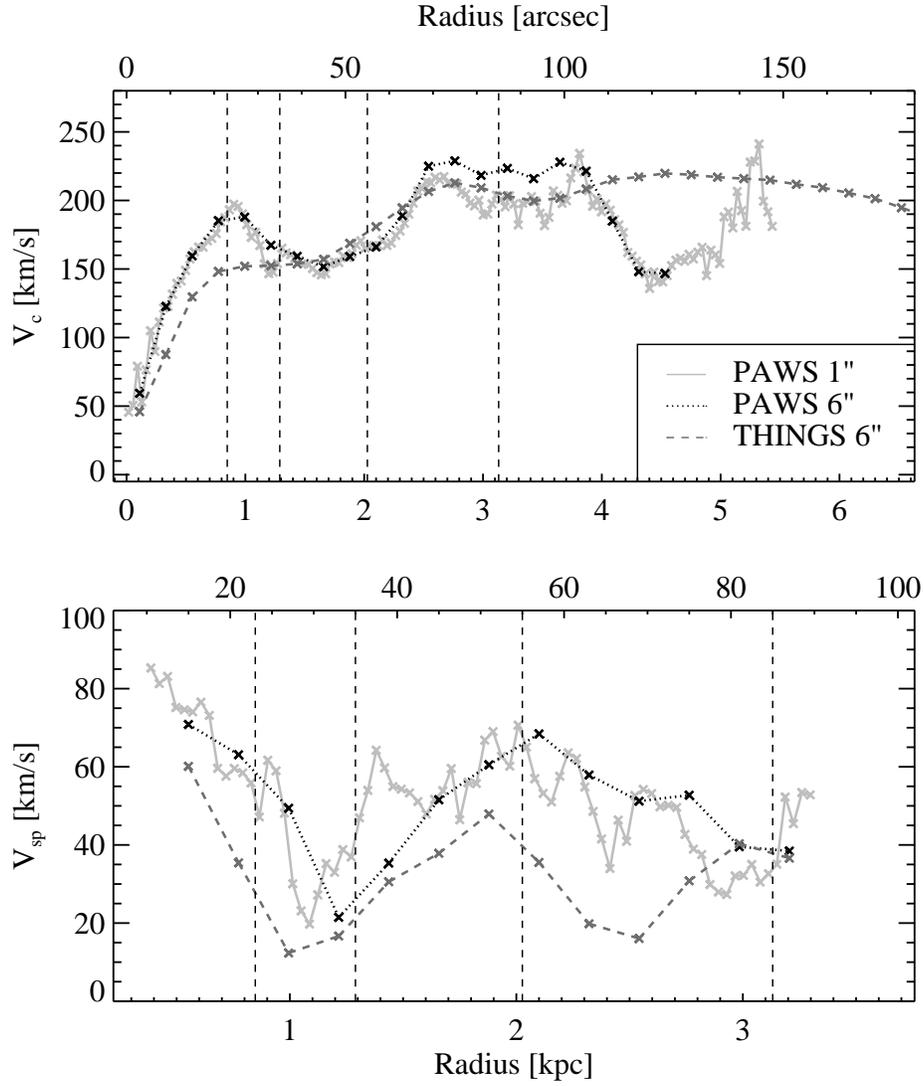}
\end{center}
\caption{\footnotesize Rotation curves (top) and amplitude of the two-armed spiral arm perturbation (bottom) 
derived from PAWS 1", PAWS 6" and THINGS 6" derived with the GIPSY task ROTCUR and the method described in Section~\ref{sec:non_circ}, respectively. Vertical dashed lines represent M51's environments as defined in \cite{meidt13}.
The dip at $R\sim100-150"$ in the PAWS 1" rotation curve is probably caused by the low inter-arm sampling due to the rectangular shape of the PAWS FoV that leads the fit to favor the spiral arms.}
\label{fig:vsp_pt6as}
\end{figure*}

\begin{figure*}[!h]
\begin{center}
\includegraphics[width=1\textwidth]{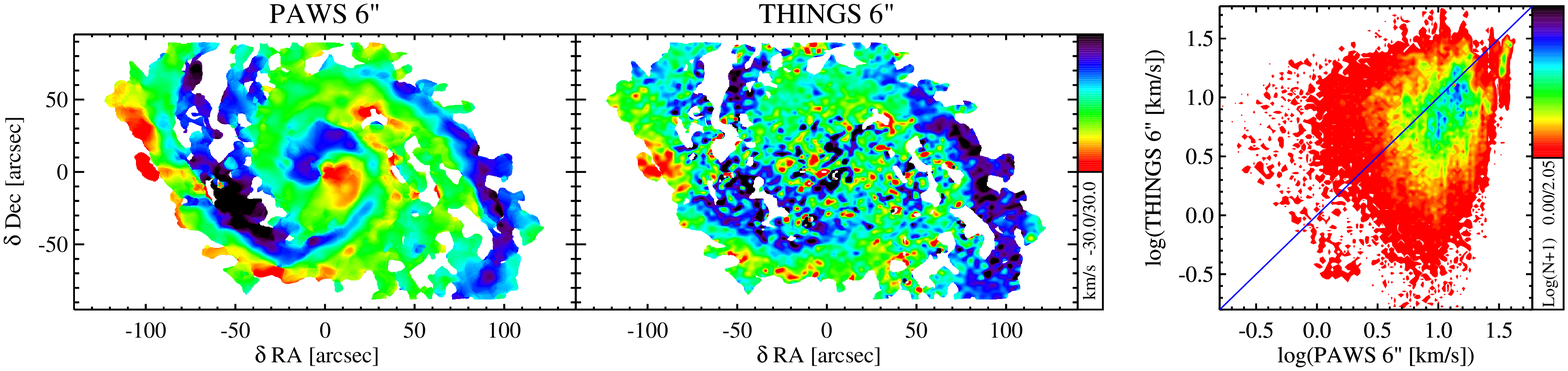}
\includegraphics[width=1\textwidth]{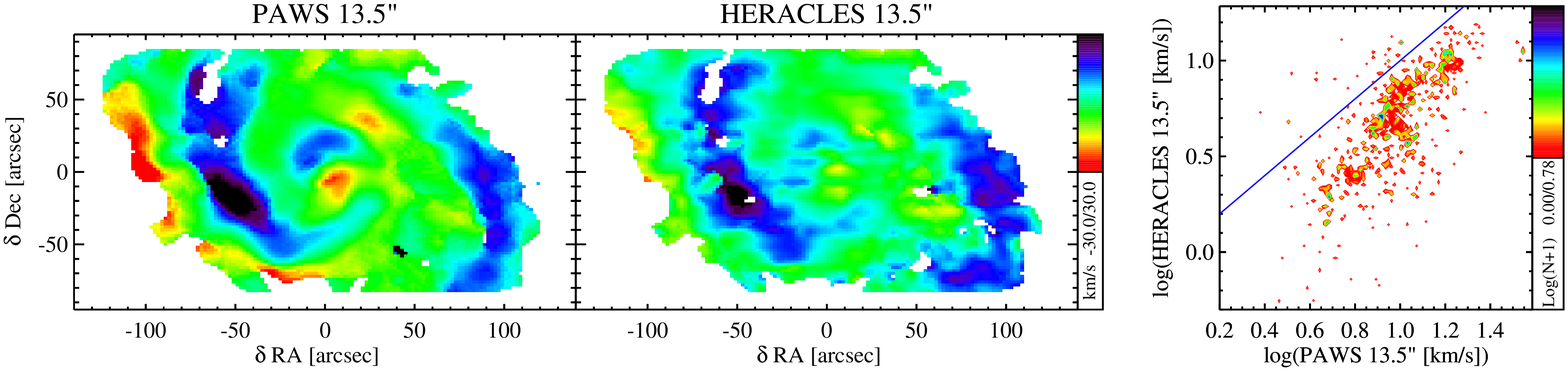}
\includegraphics[width=1\textwidth]{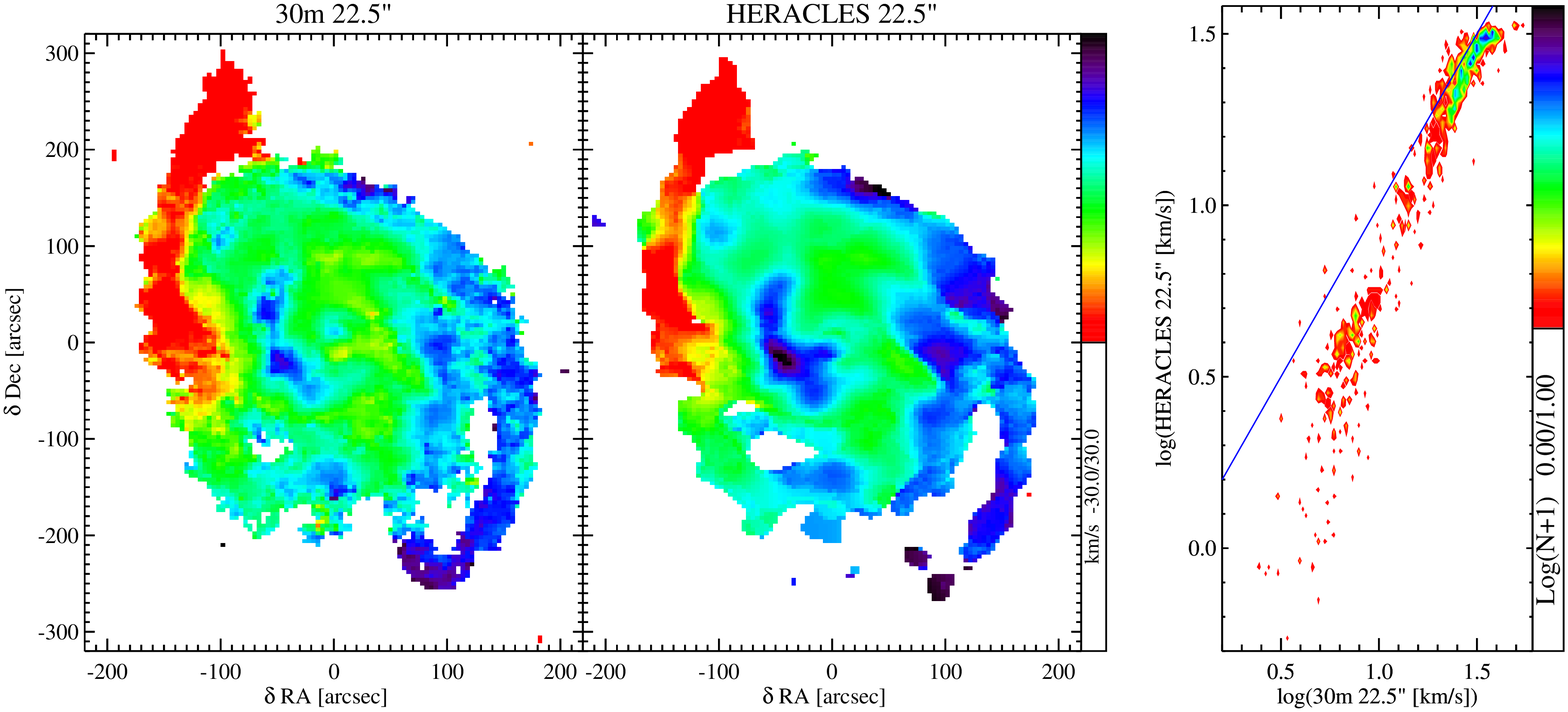}
\end{center}
\caption{\footnotesize Comparison between PAWS 6" and THINGS 6" (top), PAWS 6" smoothed to 13.5" and
HERACLES, PAWS single dish and HERACLES smoothed to 22.5" (bottom) residual velocity fields, on
the same pixel size and FoV. The operations are performed using the 
MIRIAD tasks CONVOL and REGRID on the datacubes. Residual velocity fields are obtained using the procedure described in Section~\ref{sec:res_vel}. The right panels show the pixel-by-pixel comparisons of the residual velocity fields in value of the pixels. Number densities of the points are in logarithmic scale. Blue lines indicate the 1:1 relation.}
\label{fig:kin_comparison}
\end{figure*}

\noindent We have considered whether the difference between hybrid PAWS and HERACLES at 13.5" resolution arises from
the fact that the two observations sample two different tracers of the molecular gas: while PAWS traces $^{12}$CO(1-0) emission, HERACLES traces $^{12}$CO(2-1). In the
last row of Fig~\ref{fig:vsp_pt6as}, we compare the residual velocity fields from the PAWS
single-dish data with HERACLES observations, smoothed to the same 22.5" resolution.  Since both
observations have been obtained with the same 
instrument (IRAM 30m antenna), instrumental effects should be negligible.  These maps show only
small differences, and the scatter in the pixel-by-pixel comparison is very low. We conclude that,
from a kinematic point of view, single-dish observations of $^{12}$CO(1-0) and $^{12}$CO(2-1)
provide similar results.\\

%\newpage

%\clearpage
%\newpage

%\clearpage
%\newpage

\section{Summary}\label{sec:summary}
\noindent In this paper we performed a detailed kinematic analysis of the inner disk of M51 with the
aim of characterizing and quantifying the non-circular motions driven in response to the bar and
spiral patterns present in the disk.  Our primary focus is the view of gas motions presented by the
high resolution PAWS 1" $^{12}$CO(1-0) data set, in addition we support the interpretation of our
findings with other lower
resolution datasets (PAWS 3" and 6" $^{12}$CO(1-0), THINGS 6" HI, HERACLES 13.5" $^{12}$CO(2-1) and
PAWS single dish 22.5" $^{12}$CO(1-0)).
Our main results are summarized as follows: 

\begin{itemize}

\item By applying a tilted ring analysis to the different velocity fields, we obtained updated
estimates of projection parameters of M51, namely position angle $PA=(173\pm3)^{\circ}$ and
inclination $i=(22\pm5)^{\circ}$.  We use these to fit for the circular velocity in each of the data
sets.

\item We perform a harmonic decomposition of the residual velocity fields in order to identify,
separate and inspect the contributions of the different modes to the global pattern of non-circular
motions in the galaxy. The
residual velocity field of M51 is complex, but shows the clear signature of arm-driven inflow
(especially along the southern arm) and the butterfly pattern of the inner bar.

\begin{itemize}

\item The dominant $m=2$ mode is characterized by a corotation radius at
$R_{CR,m=2}\approx2.4$ kpc ($R_{CR,m=2}\approx60"$), consistent with location of the corotation of
the two-armed spiral indicated by the gravitational torque analysis of \cite{meidt13}. 

\item Coincident with this mode, we find the first unequivocal evidence for an $m=3$ mode in the
inner disk of M51, extending out to $R\approx1.7$ kpc ($R\approx45"$). The kinematic signature of
this mode allows us to estimate the location of its corotation radius $R_{CR,m=3}\approx1.1\pm0.1$
kpc ($R_{CR,m=3}\approx30\pm3"$). 

\item Inspection of the angular frequency curves suggests that the $m=3$ mode may be coupled to, and
stimulated by, the nuclear bar.  Evidence for the dynamical coupling between the  three-armed spiral
and the main two-fold pattern at the overlap of their resonances is suggested by the appearance of
$m=1$ and $m=5$ components in the CO surface brightness around the overlap.  This supports the
density-wave nature of the three-armed perturbation to the potential traced by the gas motions. 

\end{itemize}

\item Combining the amplitudes of the individual harmonic components, we obtained a simple
expression for the streaming motion amplitude of the main modes in M51.    
 
The streaming motions from the main $m$=2 mode range from $\langle V_{sp,m=2}\rangle\approx70$
km s$^{-1}$ in spiral arm region devoid of star formation to $\langle V_{sp,m=2}\rangle\approx50$ km s$^{-1}$ in
the outer density-wave spiral arms, and exhibit a minimum $\langle V_{sp,m=2}\rangle\approx25$ km
s$^{-1}$ in the molecular ring region.

The streaming motion from the secondary modes ($m=1,3$) are $V_{sp,m=3}\lesssim30$ km
s$^{-1}$ in the region influenced by the $m=3$ mode and $\langle V_{sp,m=1}\rangle\approx32$ km
s$^{-1}$ in the region dominated by the
$m=1$ mode, but no higher than
$V_{sp,m=1,3}\approx20$ km s$^{-1}$ in the bar region.   

\item The joint analysis of velocity fields obtained from different gas tracers at different
resolutions suggests the following guidelines for defining the most appropriate observing strategy
to meet a given scientific goal:

\begin{itemize}

 \item high resolution CO surveys are particularly well-suited for detailed studies of non-circular
motion features, while low resolution observations are equally as important for defining the bulk
motion of the galaxies
(i.e. rotation curves).   In the presence of modes that extend over only a limited radial range, as
in M51, and when complex, overlapping structure exists generally, high resolution is key to
identifying and characterizing such modes.  

 \item CO and HI can supply independent views of the gravitational potential, as suggested by
different natures of the two gas phases; while the atomic gas in M51 has a smooth distribution,
is located mostly downstream of the spiral arms and in a thicker disk, 
the molecular gas is more compact, organized in a thinner disk and mostly confined
to the spiral arms.  Given the differences in velocity dispersion and morphology, we conclude that
CO is optimal for tracing spiral arm streaming motions and, in general, for studying the galactic
potential, while HI is more suitable for obtaining the bulk motion and the projection parameters of
the galaxies.   

\end{itemize}

\end{itemize}

%%%%%%%%%%%%%%%%%%%%%%%%%%%%%%%%%%%%%%%%%%%%%%%%%%%%%%%%%%%%%%%%%
%%%%%%% Acknowledgements

\acknowledgements
\noindent We thank our anonymous referee for the thoughtful comments that improved the quality of the paper. We thank the IRAM staff for their support during the observations with
the Plateau de Bure interferometer and the 30m telescope.
DC and AH acknowledge funding from the Deutsche Forschungsgemeinschaft (DFG) via grant SCHI 536/5-1
and SCHI 536/7-1 as part of the priority program SPP 1573 'ISM-SPP: Physics of the Interstellar
Medium'.
CLD acknowledges funding from the European Research Council for the FP7 ERC starting grant project
LOCALSTAR.
TAT acknowledges support from NASA grant \#NNX10AD01G.
During this work, J.~Pety was partially funded by the grant ANR-09-BLAN-0231-01 from the French {\it
Agence Nationale de la Recherche} as part of the SCHISM project (\url{http://schism.ens.fr/}).
ES, AH and DC thank NRAO for their support and hospitality during their visits in Charlottesville.
ES thanks the Aspen Center for Physics and the NSF Grant \#1066293 for hospitality during the
development and writing of this paper. DC thanks Glenn van de Ven for the useful discussion and the help with the harmonic decomposition code. SGB acknowledges economic support from Junta de Andalucia grant P08 TIC 03531. The National Radio Astronomy Observatory is a facility of the National Science Foundation operated under cooperative agreement by Associated Universities, Inc.

%%%%%%%%%%%%%%%%%%%%%%%%%%%%%%%%%%%%%%%%%%%%%%%%%%%%%%%%%%%%%%%%%
%%%%%%% References

\newpage

% Appendices

\appendix

\section{Low-resolution velocity field harmonic decomposition and amplitude of spiral perturbations}

\begin{figure*}[!h]
\begin{center}
\includegraphics[width=0.5\textwidth]{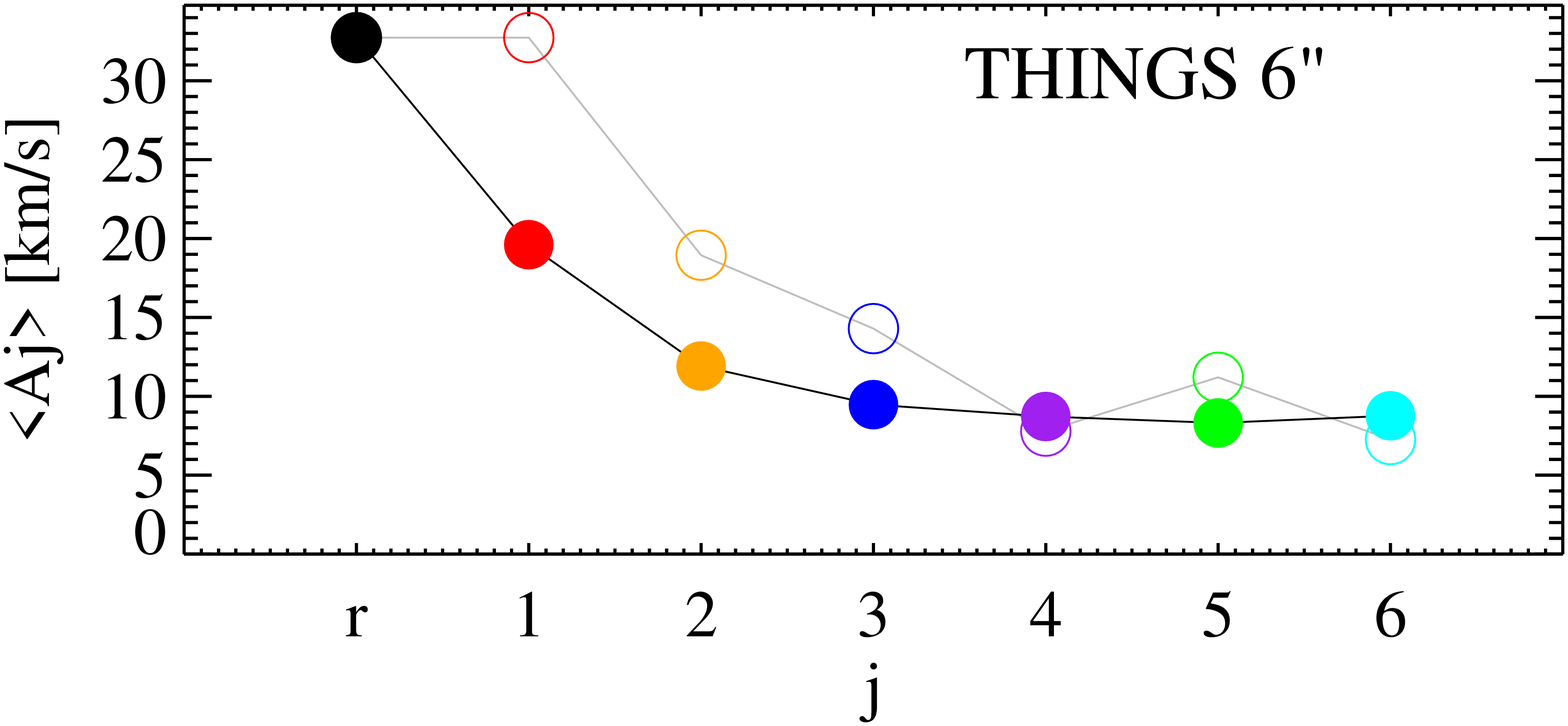}
\includegraphics[width=1\textwidth]{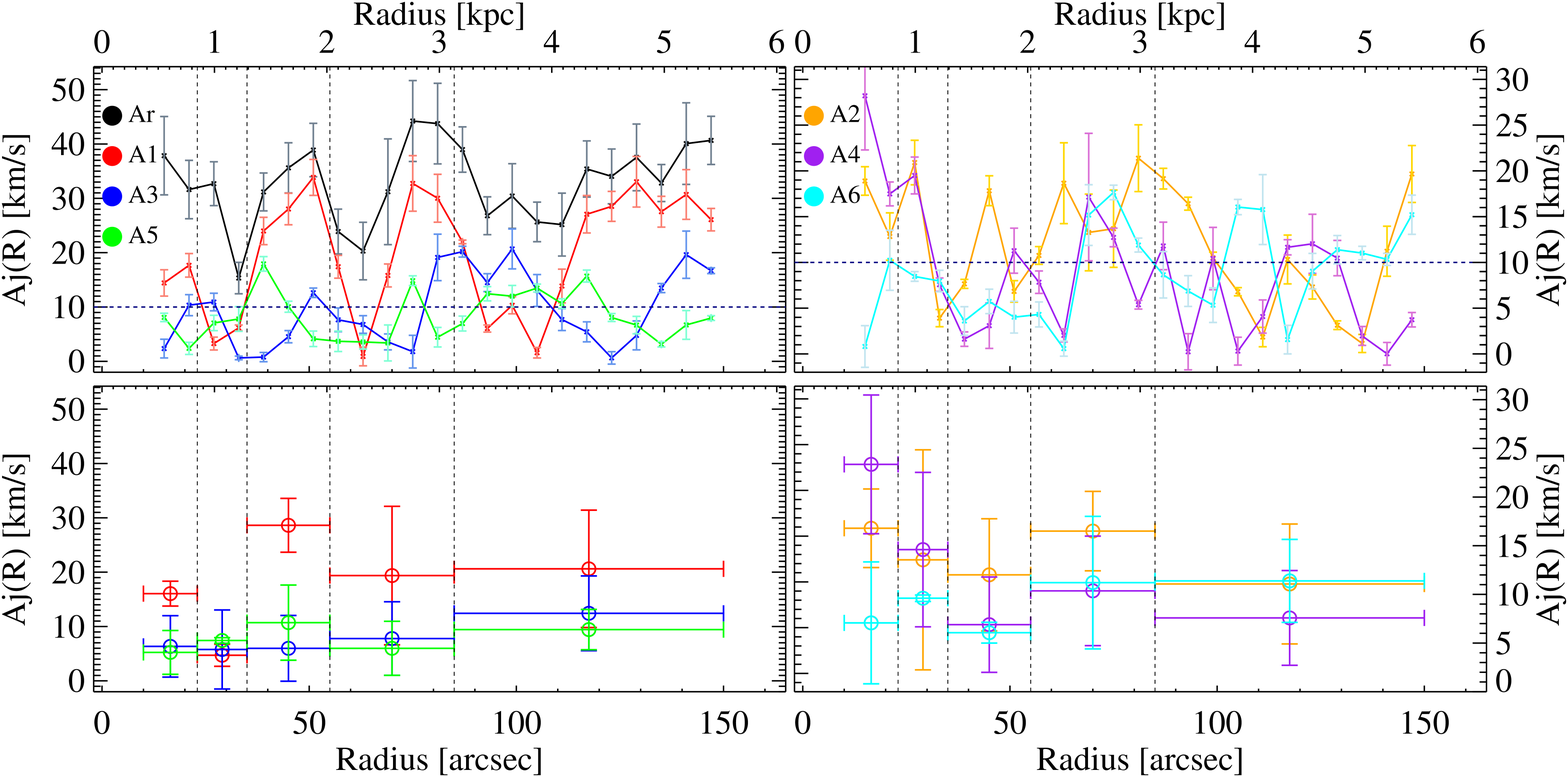}
\end{center}
\caption{\footnotesize \emph{Top plot:} Radially averaged mean of the harmonic component amplitudes
$A_{j}$
from THINGS 6" residual velocity field. Open dots indicate the measurements restricted on the PAWS FoV.
\emph{Bottom plot:} Non-circular motion amplitudes from harmonic decomposition: radial
trend of the odd components and the total power $A_{r}(R)$ (top left) and even components
(top right). The horizontal blue dashed straight line indicates twice the channel width of the datacube, i.e. $2\times5$ km\,s$^{-1}$ = 10 km\,s$^{-1}$. In the bottom row
the mean behavior of the odd (left) and even (middle) components in the different M51 environments
 as defined in \cite{meidt13} (dashed vertical lines; see the text for details) are indicated together with the standard deviations of the values. Horizontal error bars represents the widths of the environments.}
\label{fig:ampl_things}
\end{figure*}

\begin{figure*}
\begin{center}
\includegraphics[width=0.5\textwidth]{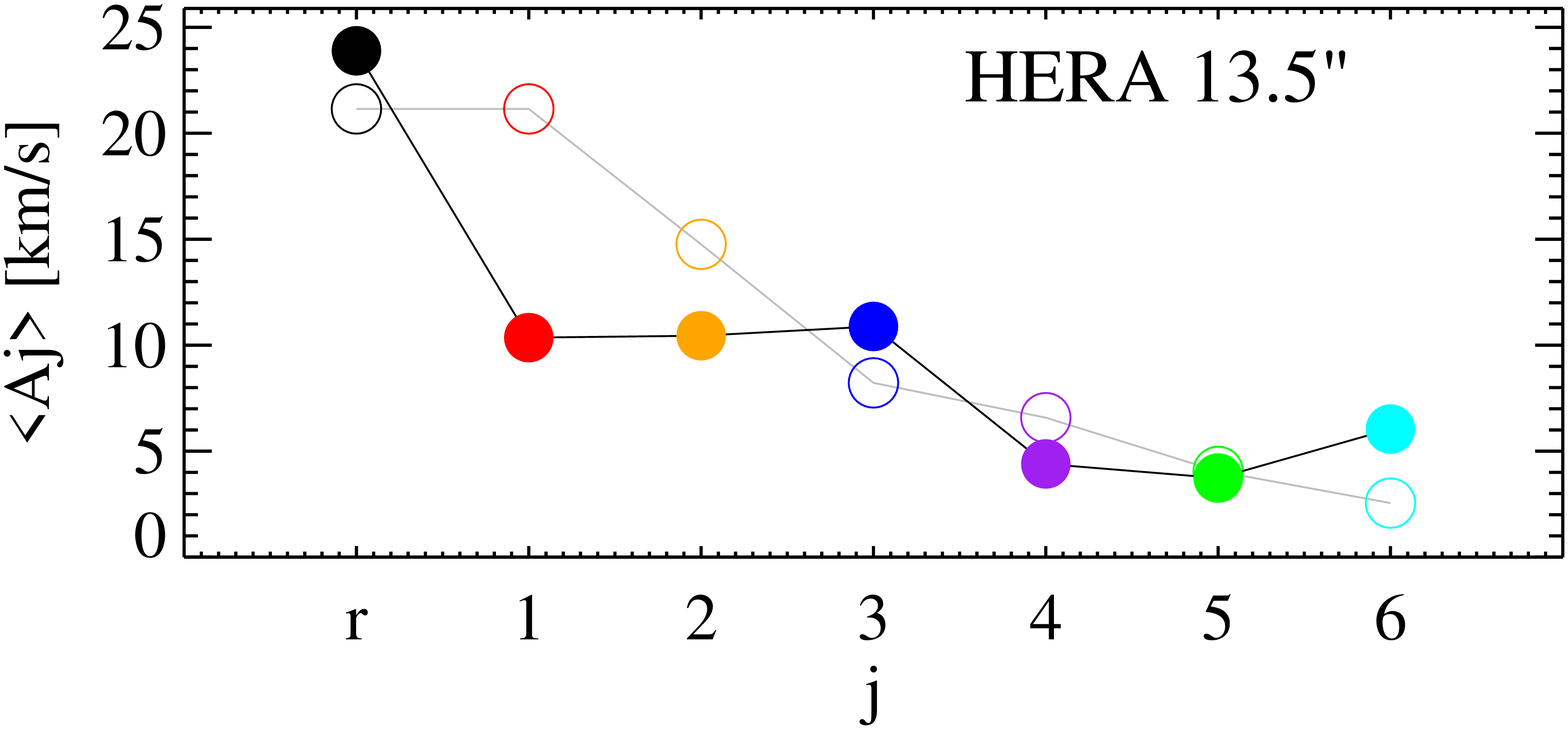}
\includegraphics[width=1\textwidth]{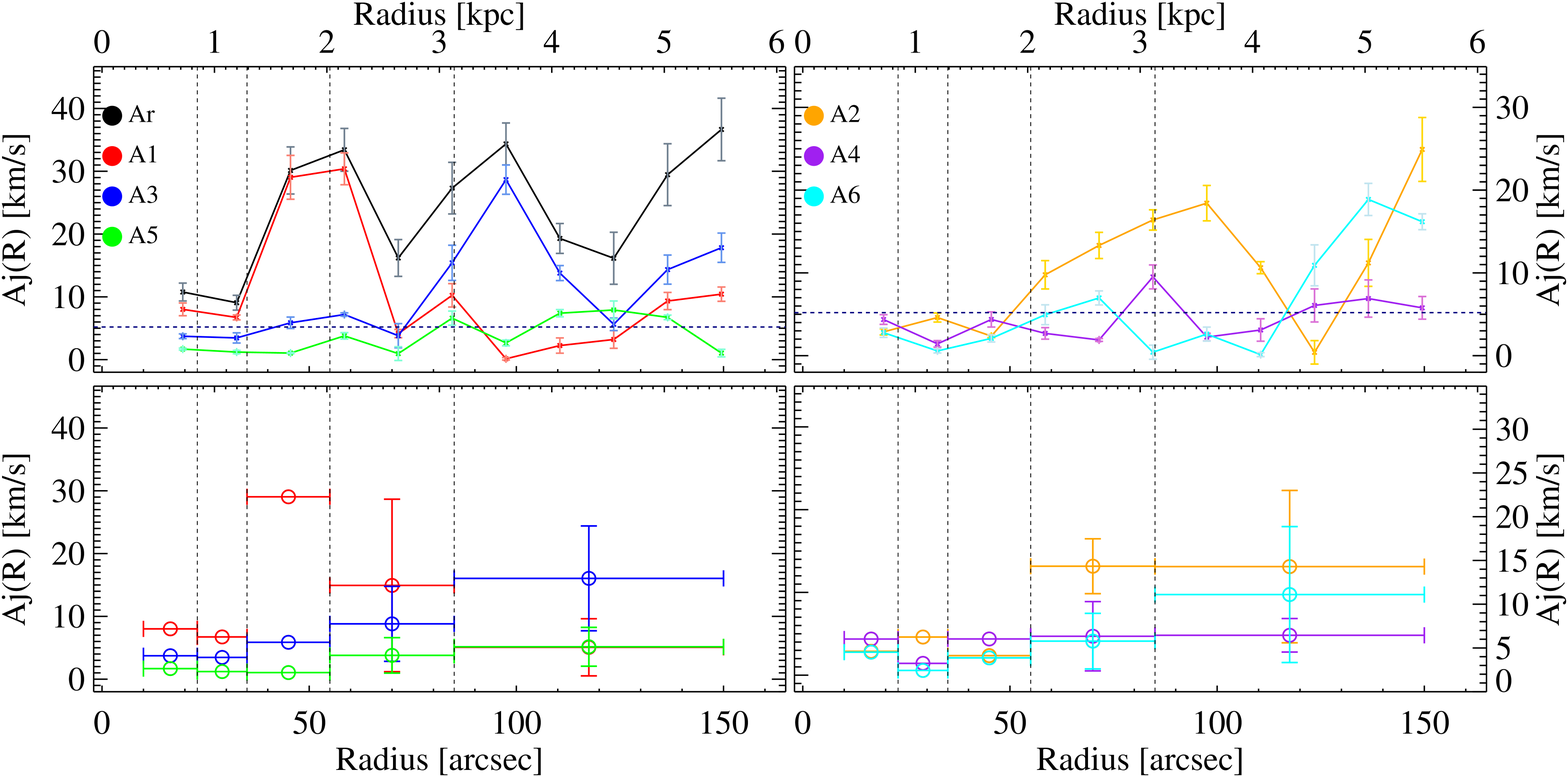}
\end{center}
\caption{\footnotesize \emph{Top plot:} \emph{Top plot:} Radially averaged mean of the harmonic component amplitudes
$A_{j}$
from HERACLES 13.5" residual velocity field. Open dots indicate the measurements restricted on the PAWS FoV.
\emph{Bottom plot:} Non-circular motion amplitudes from harmonic decomposition: radial
trend of the odd components and the total power $A_{r}(R)$ (top left) and even components
(top right). The horizontal blue dashed straight line indicates twice the channel width of the datacube, i.e. $2\times2.6$ km\,s$^{-1}$ = 5.2 km\,s$^{-1}$. In the bottom row
the mean behavior of the odd (left) and even (middle) components in the different M51 environments
 as defined in \cite{meidt13} (dashed vertical lines; see the text for details) are indicated together with the standard deviations of the values. Horizontal error bars represents the widths of the environments.}
\label{fig:ampl_hera}
\end{figure*}

\begin{figure}
\begin{center}
\includegraphics[width=1\textwidth]{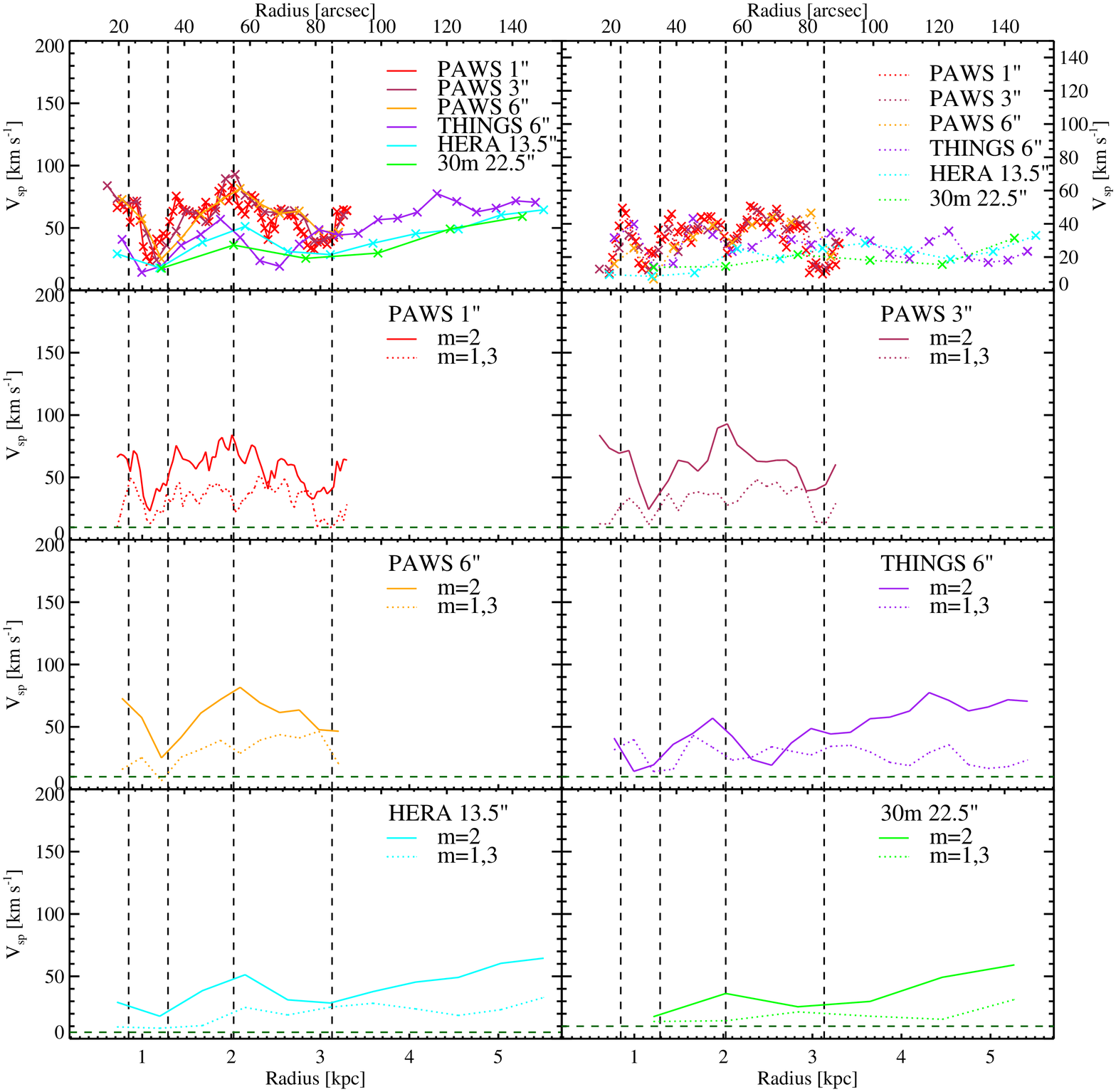}
\end{center}
\caption{\footnotesize Amplitude of the spiral perturbations from PAWS 1", PAWS 3", THINGS 6",
HERACLES 13", 30m. Solid lines indicate the streaming motion induced by the $m=2$ mode,
while the dashed line the streaming motion from the $m=1,3$ mode. 
The top left panel gives the compact view of the pattern speed derived from the different residual
maps given by $m=2$ (left) and $m=1,3$ modes (right).}
\label{fig:vsp_all}
\end{figure}

\end{document}